\documentclass[10pt]{article}
\usepackage[margin=0.8 in]{geometry}    
\usepackage[title,titletoc,toc]{appendix}
\usepackage{hyperref}     
 \usepackage{color}
 \usepackage[most]{tcolorbox}
 \tcbset{colback=yellow!10!white, colframe=red!50!black, 
        highlight math style= {enhanced, 
            colframe=red,colback=red!10!white,boxsep=0pt}
        }        
\usepackage{graphicx}
\usepackage{amssymb}
\usepackage{amsmath}
\usepackage{amsfonts}
\usepackage{epstopdf}

\newcommand{\p}{\partial}

\newcommand{\lo}{\Lambda_0}
\newcommand{\lm}{\Lambda}
\newcommand{\hf}{{1\over 2}}
\newcommand{\be}{\begin{equation}}
\newcommand{\br}{\begin{eqnarray}}
\newcommand{\er}{\end{eqnarray}}
\newcommand{\ee}{\end{equation}}
\newcommand{\bt}{\begin{tabular}}
\newcommand{\et}{\end{tabular}}
\newcommand{\bc}{\begin{tcolorbox}}
\newcommand{\ec}{\end{tcolorbox}}

\newcommand{\dd}{\delta}
\newcommand{\DD}{\Delta}

\newcommand{\Dp}{\frac{d^Dp}{(2\pi)^D}}

\newcommand{\idpl}{\int {\cal D}\phi_l}
\newcommand{\idph}{\int {\cal D}\phi_h}

\newcommand{\eps}{\epsilon}

\newcommand{\Dt}{\frac{D}{2}}

\title{Holographic RG and Exact RG in O(N) Model}
\author{ B. Sathiapalan\\Institute of Mathematical Sciences\\
CIT Campus, Tharamani\\ Chennai 600113, India\\
and\\ Homi Bhabha National Institute\\Training School Complex, Anushakti Nagar\\
Mumbai 400085, India\\bala@imsc.res.in}

\begin{document}

{\let\newpage\relax\maketitle}

\maketitle
\begin{abstract}
 In this paper an Exact Renormalization Group (ERG) equation is written for the the critical $O(N)$ model in $D$-dimensions (with $D\approx 3$) at the Wilson-Fisher fixed point perturbed by a scalar composite operator. The action  is written in terms of an auxiliary scalar field
 and reproduces correlation functions of a scalar composite operator.
  The equation is derived starting from the Polchinski ERG equation for the fundamental scalar field. As described in arXiv:1706.03371 (\cite{Sathiapalan:2017})  an evolution operator for the Polchinski ERG equation can be written in the form of a functional integral, with  a $D+1$ dimensional scalar field theory action.  In the case of the fundamental scalar field this action only has a kinetic term and therefore  looks quite different from Holographic RG where there are potential terms. But in the composite operator case discussed in this paper, the ERG equation and consequently the $D+1$ dimensional action contains higher order potential terms for the scalar field and is therefore very similar to  the case of Holographic RG. Furthermore  this action can be mapped to a scalar field action in $AdS_{D+1}$ using the techniques of (\cite{Sathiapalan:2017}). The leading  cubic term of the potential is computed in this paper for $D \approx 3$ and expectedly vanishes in $D=3$ in agreement with results in the AdS/CFT literature.

\end{abstract}
\newpage
\tableofcontents
\newpage

\section{Introduction}

One of the most interesting ideas to come out of string theory is the AdS/CFT correspondence \cite{Maldacena,Polyakov,Witten1,Witten2} where a boundary CFT is dual to a bulk string theory. \footnote{See \cite{Penedones2016} for a review and references.}   This can be viewed as a realization of the older idea of holography where it is introduced as a property of  gravity \cite{tHooft:1993dmi,Susskind:1994vu}.

In  \cite{Sathiapalan:2017} it was shown that there is a way to obtain a bulk dual starting from the Exact Renormalization Group (ERG) \cite{Wilson,Wegner,Wilson2,Polchinski} of the boundary theory. (There are  other  more recent versions of the ERG equations and  many reviews \cite{Wetterich,MorrisERG,Bagnuls1,Bagnuls2,Igarashi,Rosten:2010}.)    The Polchinski ERG equation \cite{Polchinski} for a boundary $D$-dimensional scalar field theory is our starting point. This has an evolution operator that can be written as a functional integral with a $D+1$ dimensional scalar field action. The action consists of a term that looks like a kinetic term of a field theory, but with a time dependent prefactor. It was then shown that by a suitable field redefinition this action can be mapped to a standard field theory action in $AdS_{D+1}$. Thus in this approach the $D+1$ dimensional ``bulk" theory is an RG evolution operator. The scale of the boundary theory becomes a geometrical radial coordinate of AdS space and is the direction of evolution. 

 At this point one makes contact with ``Holographic RG" [\cite{Akhmedov}-\cite{deHaro:2000}]: In the AdS/CFT correspondence, the radial coordinate is interpreted as a scale for the boundary theory, and moving the location of the boundary is equivalent to a scale transformation. Then the AdS/CFT correspondence would imply that the bulk action corresponds to an evolution operator for the RG evolution of the boundary theory. In the construction of \cite{Sathiapalan:2017}, the starting point is an ERG equation of the boundary theory and one is able to {\em derive} holographic RG and  a dual bulk theory in AdS space. One does not have to resort to  the AdS/CFT correspondence hypothesis to obtain Holographic RG equations.
This is suggestive of the possibility mentioned above that the AdS/CFT correspondence can be derived from Renormalization Group concepts \cite{SSLee:2010,SSLee:2012,Meloa:2019}. 

The construction in  \cite{Sathiapalan:2017} was restricted to free field theory. In \cite{Sathiapalan:2019} it was generalized to free theories of fields with non canonical scaling dimensions.
This is a first step towards interacting CFT's where one encounters anomalous dimension due to interactions. However the issue of interactions was left open. 

In this paper we discuss the issue of interaction. We study the critical $O(N)$ model in $D$-dimensions with $D\approx 3$ (see \cite{Zinn-Justin,Moshe:2003} for a review and references), and using auxiliary fields, an action is derived that describes correlation 
functions of composite scalar operators
of the form $\phi^I\phi^I$.  Starting from Polchinski's ERG equation for the fundamental scalar field $\phi^I$, an ERG equation is derived for the generating functional written in terms of auxiliary fields. It is shown that the leading term is of the Polchinski form but there are, in addition, potential terms involving cubic and higher order vertices. This equation is one of the main results of this paper. The potential terms in the equation become potential terms in the bulk action that defines the ERG evolution operator.
 This action can be mapped to AdS space by the same kind of field redefinition that was introduced in  \cite{Sathiapalan:2017}. The end result is a bulk action for a scalar field {\em with potential terms} of the type encountered in Holographic duals of the $O(N)$ model. We compute the leading cubic term. The cubic correlation function computed using this interaction term reproduces the three point correlators that have been computed the literature on the subject \cite{Giombi:2009,Sezgin:2003,Petkou:2003,Leonhardt:2002} and in particular it says that this cubic interaction term in the bulk and the three point correlator vanishes in D=3.  This is also expected from the conjectured bulk dual \cite{Klebanov:2002,Sezgin:2002} which is a higher spin theory \cite{Vasiliev:2003,Vasiliev:2004}.

In this approach the bulk theory in AdS space is the ERG evolution operator and therefore correlators
calculated using the bulk theory are guaranteed to agree with calculations done in the boundary. So what is
obtained in this approach can be described as a Holographic rewriting of ERG equations in AdS space. The
interesting and as yet open question then is whether the bulk theory so obtained is the same as the dual theories
that have been identified in AdS/CFT literature.

This paper is organised as follows: In Section 2 we give some background about the $O(N )$ model. In Section
3 we derive the ERG equation for this model. In Section 4 an evolution operator for the ERG equation is given
and some correlation functions computed for the $O(N )$ model. This constitutes a holographic description of
the theory. In Section 5 we map the action to AdS space by a field redefiniton and again give a computation
of the cubic term and some correlation functions. Section 6 gives a summary and conclusions.

\section{Review}
\label{rev}
In this section we give a brief review of the results of  \cite{Sathiapalan:2017} where an action for a free scalar field in $AdS_{D+1}$ spacetime was obtained, representing an RG evolution operator for a perturbed boundary CFT in $D$ space time dimensions, as in the case of AdS/CFT correspondence. There are three steps involved and we review them here.
\begin{enumerate}
\item {\bf Step 1: ERG Equation}

The starting point is Polchinski's ERG equation for the Wilson action (or equivalently for the Generating Functional) of a {\em fundamental} scalar field.  We emphasize the word ``fundamental" because in this paper we will derive a corresponding ERG equation for the Generating Functional of field representing a composite operator.

We start with a scalar field (bare) theory defined at the scale $\lo$. \footnote{$\lo\to \infty$ is the continuum limit. We will often take this limit to simplify calculations.}
\be
S_B= \int \Dp [\hf \phi(p) \frac{p^2}{K(\lo)}\phi(-p)] + S_{B,I}[\phi]
\ee
$K(\lm)$ is a cutoff function that allows popagation of modes below $\lm$. It can be an analytic function of the form 
\[
K(\lm)=e^{-\frac{p^2}{\lm^2}}
\]

We integrate out modes between $\lo$ and $\lm$ and obtain a low energy Wilson action at the scale $\lm$. This is done by introducing low and high energy modes satisfying
\be
\phi(p)=\phi_l(p)+\phi_h(p)~~~;~~~\frac{1}{p^2}\equiv\DD = \DD_l+\DD_h
\ee
We can take 
\be
\DD_l=\frac{K(\lm)}{p^2},~~~\DD_h=\frac{K(\lo)-K(\lm)}{p^2}
\ee
The original functional integral with the standard kinetic term can then be
shown to be equivalent to a theory written in terms of $\phi_{l,h}$ with kinetic term:
\be
\hf\int_p \phi_l(p)\DD_l^{-1}\phi_l(-p) +\hf\int_p \phi_h(p)\DD_h^{-1}\phi_h(-p)
\ee
We will often use, for simplicity, $G=\DD_l$ below.

 The Wilson action obtained by integrating out $\phi_h$ is written as
\be
S_\lm[\phi]= \hf \int \Dp \phi (p)G^{-1}\phi(-p) +S_{ \lm,I}[\phi]
\ee
where
\be
\int {\cal D}\phi_h~e^{-\hf \int _p \phi_h(p)\DD_h^{-1}\phi_h(-p)-S_{B,I}[\phi_l+\phi_h]}\equiv e^{-S_{\lm,I}[\phi_l]}
\ee

 It can be shown to obey Polchinski's ERG equation:
\be  \label{polch}
\frac{\p}{\p t} e^{-S_{\lm,I}[\phi]}= -\hf \int_p\dot G(p)\frac{\dd ^2}{\dd \phi(p)\dd\phi(-p)}e^{-S_{\lm,I}[\phi]}
\ee

It can also be shown that the generating functional $Z[J]$ obeys a very similar equation:
\be  \label{polch2}
\frac{\p}{\p t} e^{W_\lm[J]}= -\hf \int_p {(\frac{\dot \DD_h}{\DD_h^2})}\frac{\dd ^2}{\dd J(p)\dd J(-p)}e^{W_\lm[J]}
\ee

\item {\bf Step 2: Holographic Form}

One can write a solution in the form of a functional integral to this equation by analogy with the Feynman path integral solution to the non relativistic Schroedinger equation \cite{Sathiapalan:2017}.

Defining $\psi[\phi,t]= e^{-S_{\lm,I}}$ with $\lo = \lm e^t$, we have
\be
\psi[\phi_f,t_f]=\int {\cal D}\phi_i(p) \int_{\phi(p,t_i)=\phi_i(p)} {\cal D} \phi(p,t)~e^{-\hf \int dt~\int_p {\dot G}^{-1}{\dot \phi}(p,t){\dot \phi}(-p,t)}\psi[\phi_i(p),t_i]
\ee

Thus starting with a $D$-dimensional field theory, we obtain a $D+1$ dimensional field theory as the ERG evolution operator. This is a holographic form of the theory. 

\item {\bf Step 3: Mapping to $AdS_{D+1}$ spacetime}

The $AdS_{D+1}$ metric is taken to be 
\be
ds^2=\frac{dz^2+dx_i dx^i}{z^2}
\ee

In \cite{Sathiapalan:2017} it was shown that the bulk scalar field action above can be mapped to a standard $AdS_{D+1}$ free scalar field theory by a field redefinition:

Writing $z=\lm^{-1}$ instead of $t$, define an AdS scalar field $y(p,z)$
\be   \label{y}
\phi(p,z)= f(p,z)y(p,z)
\ee
 $f$ is defined by 
 \be
 f^2=-z^{-D}\dot G
 \ee and is chosen to satisfy:
\be
[\frac{d^2}{dz^2}+\frac{1}{z}\frac{d}{dz} -(p^2+\frac{m^2}{z^2})](\frac{1}{f(p,z)})=0
\ee
Then the scalar field action becomes \footnote{There is also a boundary term that is not important.}
\be   \label{ba}
S[y(p)]=\int dz~\int_p z^{-D+1} [\frac{\p y(p,z)}{\p z}\frac{\p y(-p,z)}{\p z}+ (p^2+\frac{m^2}{z^2})y(p,z)y(-p,z)]
\ee

It was shown in \cite{Sathiapalan:2017} that $f,G$ are given by
\br
\frac{1}{f(p,z)}&=&z^\Dt(A(p)K_\nu (pz) +B(p)I_\nu(pz))\nonumber\\
G(p,z)&=&\frac{C(p)K_\nu (pz) +D(p)I_\nu(pz)}{A(p)K_\nu (pz) +B(p)I_\nu(pz)}
\er
with $AD-BC=1$ and $\nu^2=m^2+\frac{D^2}{4}$.

\end{enumerate}

Thus we have a D+1 dimenional AdS bulk spacetime and a scalar field action. One can do very similar manipulations for $Z[J]$ because it obeys a very similar equation. All this is exactly as in AdS/CFT - as far as free theories are concerned. 

As mentioned in the introduction, the question of interactions was left open in \cite{Sathiapalan:2017}. This is the main topic of this paper. The scalar field action \eqref{ba} has information about the kinetic term of the boundary theory in the form of the function $G=\frac{K(\lm)}{p^2}$ .  Actually, as clarified in \cite{Sathiapalan:2019}, this action also has information about the scaling dimension of the boundary field - this is captured by the free parameter $m^2$. This is to be understood as an anomalous dimension and arises by the scale dependent  field redefinition given in \eqref{y}. But information about interactions is not contained in the bulk action \eqref{ba}- which is a free theory . This is of course a reflection of the fact that Wilsonian ERG equations define a coarse graining that  does not depend on the theory -except for an anomalous dimension parameter. This would seem to make it different from the Holographic RG equations encountered in AdS/CFT correspondence that have potential terms for the scalar field. 

In the coming sections we discuss a resolution of this problem in a theory with a non trivial fixed point - the O(N) model- and study the holographic version of the ERG equation for this theory. We show that when {\em composite operators} are studied in this theory, the ERG equations are modified from the usual Wilsonian form and includes extra terms.

\section{$O(N)$ model}
\subsection{Background}
\label{Zinn}

We refer the reader to \cite{Zinn-Justin,Moshe:2003} for a review of the $O(N)$ model which is summarized in this section. We start with the action  :
\be \label{S1}
S=\int_x [ \hf \p_\mu \phi^I \p^\mu \phi^I +\frac{u}{4!} (\phi^I\phi^I)^2 + \hf r \phi^I \phi^I]
\ee
Rescale $\phi^I= N^{1/4}\phi'^I$ and drop the prime, set $\frac{uN}{12}=\bar u$ to get:
\be \label{S2}
S=\int_x [ \sqrt{N}(\hf \p_\mu \phi^I \p^\mu \phi^I + \hf r \phi^I \phi^I)+\frac{\bar u}{2} (\phi^I\phi^I)^2 ]
\ee

We introduce auxiliary fields, $\chi$ a Lagrange multiplier\footnote{The integration contour for $\chi$ has to be taken along the imaginary axis.}, and $\sigma$ and rewrite the action   as follows:
\[
S=\int _x [\sqrt{N}(\hf \p _\mu \phi^I \p^\mu \phi^I +\hf r \sigma) +\frac{\bar u}{2} \sigma^2 +  \chi(\sigma - \phi^I\phi^I)]
\]
Note that with this normalization $\langle \sigma (x) \sigma(0) \rangle =\langle \phi^2(x)\phi^2(0)\rangle \approx O(1)$.
Define
\[
Z[J]= \int {\cal D}\chi  {\cal D}\sigma  {\cal D}\phi^I~e^{-S+\int _x J \chi}
\]
Do $\sigma$ integral
\[
 \int {\cal D}\sigma e^{-\frac{\bar u}{2} \sigma^2 -\hf \sqrt{N}r \sigma -\chi \sigma + J \chi}= e^{\frac{1}{2\bar u} (\chi+\frac{r\sqrt{N}}{2})^2 +\int J\chi}
  \]

\be  \label{z0}
Z[0]=\int  {\cal D}\chi \int  {\cal D}\phi^I e^{-[\int \hf \sqrt{N}\p_\mu \phi^I \p^\mu \phi^I -\chi \phi.\phi +\frac{1}{2\bar u}\chi^2 +\frac{\sqrt N}{2\bar u} \chi r] }
\ee
(Dropping field independent constants.)


The $\chi$ equation of motion is
\be   \label{eomchi}
\chi=\bar u \phi.\phi -\hf \sqrt Nr
\ee
So one can obtain correlators of $\phi^2$ from those of $\chi$. Note also the dimensions of the various fields in $D$ dimensions with $D<4$:
\[
[\phi^I]=\frac{D-2}{2},~~~[\sigma]= D-2,~~~~[\chi]=2
\]
This also means that $\frac{\chi^2}{\bar u}$ is irrelevant and can be ignored in the IR limit. Thus the Lagrange multiplier imposes $\phi.\phi=\hf \sqrt N r$. Thus this model is in the same universality class as the non linear sigma model which has a mass gap. If we set $r=0$ we get a critical theory. The condition $r=0$ is modified to $r=r_c$ by quantum corrections as we see below.


Now do $\phi$ integral:
\[
Z[J]= \int  {\cal D}\chi \int  {\cal D}\phi^I e^{-[\int \hf \sqrt{N}\p_\mu \phi^I \p^\mu \phi^I -\chi \phi.\phi +\frac{1}{2\bar u}\chi^2 +\frac{1}{2\bar u} \chi r] +\int J\chi}
\]
\be  \label{z1}
Z[J]=e^{W[J]}= \int {\cal D}\chi e^{-\frac{N}{2} Tr\ln[-\sqrt{N}\frac{\Box}{2}-\chi] -{\frac{1}{2\bar u} \int(\chi+\frac{r\sqrt{N}}{2})^2}+\int J \chi}
\ee
 
%


For large $N$ we can do a semiclassical evaluation of $Z[J]$. Or equivalently set the tadpole to zero to determine the vacuum. So we get the equation
\[
-\hf \sqrt{N} Tr \frac{1}{\frac{p^2}{2}-\chi_0} - \frac{\sqrt{N}}{\bar u}(\chi_0+r/2)=0
\]
\[
 - \frac{1}{\bar u}(\chi_0+r/2) = \int \Dp \frac{1}{p^2-2\chi_0}
 \]
 \be
- \frac{1}{\bar u}\chi_0 =\frac{1}{2\bar u}r+\int \Dp \frac{1}{p^2-2\chi_0}
\ee
Since $\chi_0>0$ corresponds to and expectation value for $\phi^2$, we must have $r$ sufficiently negative to compensate for the integral. When $\chi_0$ is just zero, we get the critical value of $r$:
\be   \label{rc}
r_c= -2\bar u\int \Dp \frac{1}{p^2}
\ee
The $mass^2$ term has to be sufficiently negative. If it starts to increase, $\chi_0$ becomes zero and then negative. This corresponds to a $mass^2$ for the $\phi$ field.

Note that for $D\leq 2$ the integral is (IR) divergent, so $\chi_0$ can never be made positive and so there is no symmetry breaking, in accordance with the Coleman-Mermin-Wagner theorem.

When $\chi_0$ is negative we think of it as a $mass^2$. Then we get an equation
\[
\chi_0 + \frac{r}{2}-\frac{r_c}{2}= -\bar u\int \Dp [\frac{1}{p^2-2\chi_0}-\frac{1}{p^2}]
\]
\be
\chi_0 + \frac{r}{2}-\frac{r_c}{2}=-\bar u\int \Dp \frac{2\chi_0}{p^2(p^2-2\chi_0)}
\ee
Let $r-r_c=\tau'$, then
\[
\chi_0 +\frac{\tau}{2} =-\bar u\int \Dp \frac{2\chi_0}{p^2(p^2-2\chi_0)} \implies 1+\frac{\tau'}{2\chi_0}= -\bar u\int \Dp \frac{1}{p^2(p^2-2\chi_0)}
\]
\be   
\frac{1}{\bar u}+ \frac{\tau'}{2\bar u\chi_0} =-\int \Dp \frac{1}{p^2(p^2-2\chi_0)}
\ee
$2\chi_0=-m^2$ is the high temperature phase. If we call $\frac{\tau'}{\bar u}=\tau$ we get \cite{Zinn-Justin,Moshe:2003}: 
\be \label{tau}
\frac{m^2}{\bar u}+ \int \Dp \frac{2m^2}{p^2(p^2+m^2)} =\tau
\ee

In Appendix \eqref{A} it is shown that the $\chi$ propagator becomes critical for a value of
$\bar u = c \lo^ {2\eps}$ where $D=4-2\eps$ and $c$ is a constant evaluated there.\footnote{In some places in this paper we use $D=3+\dd$ and reserve the symbol $\eps$ for the boundary value of the AdS radial coordinate $z=\eps$. This will be clear from the context.} In the continuum limit therefore $\bar u\to \infty$ as long as $\eps>0$. This result can also be obtained by studying the behaviour of $m^2$ using  \eqref{tau} \cite{Zinn-Justin,Moshe:2003}.

\subsection{Legendre Transform and the two fixed points}

It was argued in \cite{Klebanov:2002}, that the two fixed points $\bar u=0$ and $\bar u=\infty$ can be understood as Legendre transforms in the sense suggested by the AdS/CFT correpsondence \cite{Klebanov:1999tb}. This can be seen as follows:

We start with

\be    \label{se}
S_E=\int _x ~[\hf \sqrt N( \phi^I \Delta ^{-1} \phi^I +  r \sigma ) - \chi (\sigma -  \phi^I \phi^I) +\frac{\bar u}{2}\sigma^2 -J'\sigma]
\ee
  A source $J'$ for $\sigma$ has been added. 
\[
Z[J']=\int {\cal D} \chi \int {\cal D}\phi^I \int {\cal D}\sigma e^{-\hf \sqrt N( \phi^I \Delta ^{-1} \phi^I  + 2\chi \phi^I\phi^I] -\int (\chi+ J' - \hf\sqrt N r) \sigma -\frac{\bar u}{2}\sigma^2}
\]

Do the $\phi^I$ integral:
\[
Z[J']=\int {\cal D} \chi \int {\cal D}\sigma e^{-\frac{N}{2}\ln [\frac{\sqrt N}{2 \Delta} + \chi]- \int (\chi+ J' - \hf\sqrt N r) \sigma -\frac{\bar u}{2}\sigma^2}
\]
Do $\sigma$ integral and expand Log:
\be    \label{zj1}
Z[J']=\int {\cal D} \chi' e^{-\frac{N}{2} Tr[ -\frac{2\Delta \chi'}{\sqrt N} -\hf (\frac{2\Delta \chi'}{\sqrt N})^2 - \frac{1}{3}(\frac{2\Delta \chi'}{\sqrt N})^3 +... ]- \int \frac{(\chi+J'- \hf\sqrt N r)^2}{2\bar u}}
\ee
When $J'=0$, when we choose $r=r_c$, the linear term in $\chi$ cancels and we have:
\be
Z[0]=\int {\cal D} \chi' e^{-\underbrace{\frac{N}{2} Tr[  -\hf (\frac{2\Delta \chi'}{\sqrt N})^2 - \frac{1}{3}(\frac{2\Delta \chi'}{\sqrt N})^3 +... ]}_{S_0[\chi]}- \int \frac{\chi^2}{2\bar u}}
\ee

{ ${\boldsymbol  {\bar u}=0:}$}

Now if take $\bar u \to 0$, the Gaussian factor in \eqref{zj1} becomes a delta function and imposes  $\chi=-(J'- \hf\sqrt N r_c)$. 
In the action the linear term in $\chi$ gets a contribution only from the zero mode, so if we set $J'(0)=\hf \sqrt N r_c$ we get rid of the linear term. This is the critical theory. In the higher order terms we then have only $J(p)$ with $p\neq 0$.
This gives:
\be   \label{zj3}
Z[J']=e^{-\underbrace{\frac{N}{2} Tr[  -\hf (\frac{2\Delta J'}{\sqrt N})^2 + \frac{1}{3}(\frac{2\Delta J'}{\sqrt N})^3 +... ]}_{W[J']}}
\ee
We find that 
\[
W[J']=S_0[J']
\]

{\bf Dimensions:}

Here $[\chi(x)]=2=[J']$. ($[]$ refers to scaling dimension) $J'$ is set equal to $\chi$ so it has the same dimension.

In the original problem $J'$ is the source for $\sigma$ and $[\sigma]=[\phi.\phi]=1$. Therefore $J'$   has dimension 2 which is consistent.

\begin{figure}[h]
\includegraphics[width=18cm]{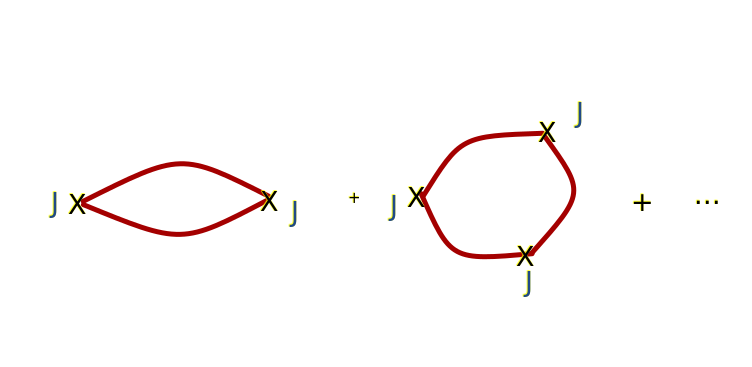}
\caption{Free theory correlations of $\phi^2$ operators. The solid lines are $\phi$ propagators. $W[J]=S_0[J]$.}
\end{figure}

{$\boldsymbol {\bar u }\to \infty$}
\be\label{zj5}
Z[J']=\int {\cal D} \chi e^{-\underbrace{\frac{N}{2} Tr[ -\frac{2\Delta \chi}{\sqrt N} -\hf (\frac{2\Delta \chi}{\sqrt N})^2 - \frac{1}{3}(\frac{2\Delta \chi}{\sqrt N})^3 +... ]}_{S_0[\chi]}- \int \frac{(\chi+J'-\hf \sqrt N r_c)^2}{2\bar u}}
\ee

Before we take the limit it is a good idea to 
rescale $J'= \bar u j$. Also $\frac{\chi^2}{2\bar u}\to 0$ in this limit. We get
\be\label{zj7}
Z'[j]=\int {\cal D} \chi e^{-\frac{N}{2} Tr[ -\frac{2\Delta \chi}{\sqrt N} -\hf (\frac{2\Delta \chi}{\sqrt N})^2 - \frac{1}{3}(\frac{2\Delta \chi}{\sqrt N})^3 +... ] +\int j\chi -\int \frac{\sqrt Nr_c\chi}{2\bar u}}
\ee
We have dropped a field independent term $\bar u j^2$. We have set $r= r_c$ so that \eqref{rc} is satisfied and the linear-in-$\chi$ term drops out. What results is the generating functional corresponding to $S_0[\chi]$: 
\be
Z'[j]=\int {\cal D} \chi e^{-S_0[\chi]+\int j\chi}=e^{W[j]}
\ee
{\bf Scaling Dimensions in D = 3 - WF fixed point:}

Now we have a conformal action for $\chi$ and a source $j$ coupled to it. So $Z[j]$ is conformal. The scaling
dimension of $j$ is thus 1 (since $[\chi] = 2$) and the operator it sources must have dimension 2. This agrees with
Polyakov’s analysis and also AdS/CFT that the scaling dimension of $\phi.\phi$ increases by 1. So in the $\bar u\to \infty$ limit
the scaling field is $\chi=\bar u \phi.\phi$ and it has scaling dimension 2. We see that engineering dimension also matches
because $\bar u$ has the dimension of mass.

Because of the factor $N$ in front of the action, loops are suppressed: We can rescale $\chi \to \sqrt N \chi$ so that $S_0[\chi]$ has no $N$ in it. 
\be\label{zj9}
Z'[j]=\int {\cal D} \chi e^{-\frac{N}{2} Tr[  -\hf (2\Delta \chi)^2 - \frac{1}{3}(2\Delta \chi)^3 +... ] +\int j\chi}=e^{W[j]}
\ee

So the $\chi$ integral contributes only tree diagrams. In other words this action $S_0[\chi]$ is the 1PI effective action in this approximation (large $N$). But this is the same as 
$W[J']$ i.e. $W[J']=S_0[J']$  of the free theory. 
 
To summarize: The functional $S_0$ is the generating functional of the free theory: $W[J']=S_0[J']$. The same functional is also the 1PI effective action $\Gamma[\chi]=S_0[\chi]$, of the infinite $\bar u$ theory. This is true only in the large $N$ limit and for a choice of rescaling $\chi$ so that $N$ dependence is  an overall factor multiplying the action. Otherwise $S_0[\chi]$ is only the leading approximation to $\Gamma[\chi]$.

{\bf Note:} In both cases we have to ensure vanishing of tadpole by tuning $r$ to $r_c$ given in \eqref{rc}.

\section{ERG}

\subsection{ERG Equation}
In this section, starting from Polchinski's ERG equation for the generating functional for a  fundamental scalar field \eqref{polch2} we will derive an ERG equation for the $\lm$ dependent generating functional of $\chi$ correlations.

Our starting point is the same as before except we define $I:1-2N$ rather than $N$ for a more convenient normalization of the equations below:
\be  \label{erg0}
Z[0]=\int  {\cal D}\chi \int  {\cal D}\phi^I e^{-[\int \hf \sqrt{N} \phi^I \DD^{-1} \phi^I -\chi \phi.\phi +\frac{1}{2\bar u}\chi^2 +\frac{\sqrt N}{2\bar u} \chi r] }
\ee

We write $\phi = \phi_h+\phi_l$ and $\DD = \DD_h +\DD_l$. Then introducing a source for $\chi$ define
\[
Z[J]=  \idpl e^{-\int \hf \sqrt N \phi_l ^I \DD_l^{-1}\phi_l^I} \int {\cal D}\chi\idph e^{-[\int \hf \sqrt N \phi_h ^I \DD_h^{-1}\phi_h^I -\chi(\phi_h+\phi_l)^2 +\frac{\chi^2}{2 \bar u} +\frac{\sqrt N}{2\bar u} \chi r +J\chi] }
\]
We also define an intermediate quantity, the Wilson Action, where $\phi_l$ is left unintegrated. This quantity thus depends on $\lm$:
\be    \label{erg1}
Z_\lm[J,\phi_l]=\int {\cal D}\chi\idph e^{-[\int \hf \sqrt N \phi_h ^I \DD_h^{-1}\phi_h^I -\chi(\phi_h+\phi_l)^2 +\frac{\chi^2}{2 \bar u} +\frac{\sqrt N}{2\bar u} \chi r +J\chi] }
\ee
\[\equiv
\int {\cal D}\chi\idph e^{-S_{B,I}[\chi,\phi_h,\phi_l,J]}\equiv e^{-S_{\lm,I}[\phi_l,J]}
\]
When high energy modes of the $\phi$ field, viz $\phi_h$, are integrated out (and also the auxiliary field $\chi$), what remains is the Wilson action as a functional of $\phi_l$ and $J$. Note also that when $\lm\to 0$,  $Z_\lm[\phi_l,J]\to Z[J]$ the full Generating Functional of $\chi$ corrrelations.  Let us do the first step of integrating out $\phi_h$:
\[
Z_\lm[J,\phi_l] = \int {\cal D}\chi \idph e^{-\int \hf  \phi_h ^I (\frac{\sqrt N}{\DD_h}+ 2\chi)\phi_h^I - \int [2\chi \phi_l^I \phi_h^I + \chi \phi_l^2 +\frac{\chi^2}{2 \bar u} +\frac{\sqrt N}{2\bar u} \chi r +J\chi] }
\]
Let us write $j^I=2 \chi \phi_l^I$. Then we have
\[  
Z_\lm[J,\phi_l]= \int  {\cal D}\chi ~e^{-N Tr \ln (\frac{\sqrt N}{\DD_h}+ 2\chi) + \hf \int_x \int_y j^I(x) \big[ \frac{1}{\frac{\sqrt N}{\DD_h}+ 2\chi}\big]_{xy}j^I(y) -\int \chi(\phi_l^2 +J+\frac{\sqrt N}{2\bar u}r) -\frac{\chi^2}{2 \bar u}}
\]
\be \label{erg2}
\equiv \int  {\cal D}\chi~e^{-S_\lm[\chi,\phi_l,J]} \equiv e^{-S_{\lm,I}[\phi_l,J]}
\ee
Unlike in \eqref{erg1}, where $\lm$ occurs only in the kinetic term, in \eqref{erg2} $\lm$ occurs not only in the kinetic term for $\chi$ but also in the interaction terms. 

We would like to think of $S_{\lm,I}[\phi_l,J]$ as the generating functional of a theory where $\chi$ is the field variable, $J$ its source,  and write an ERG for this.\footnote{If $J=0$, this is the usual interacting part of the Wilson Action.} $\phi_l$ can be set to zero if we want to simplify the functional. We are interested in correlations of $\chi$ (which stands for $\phi^2$ as the EOM \eqref{eomchi} shows) and the $J$ dependence is sufficient for this. Thus we will call $S_{\lm,I}[0,J]\equiv W_\lm[J]$ and derive an equation for $W_\lm[J]$. $W_\lm[J]$ can be used to derive (connected) correlation functions for $\chi$ when $\lm\to 0$.

We first write the Polchinski ERG equation for $Z$ using the $\phi$ representation following standard arguments, as follows:

The original Polchinski ERG is (note the factor of $1/\sqrt N$  and also that $\dot \DD_h=-\dot \DD_l$)
\[
\frac{\p }{\p t}e^{-S_{\lm,I}}=\hf \frac{1}{\sqrt N} \int_x\int_y \dot \DD _{hxy}\frac{\dd^2}{\dd \phi_l ^I(x)\dd \phi_l ^I(y)} e^{-S_{\lm,I}[\phi_l,J]}
\]
We can write
\[
\frac{\dd}{\dd \phi_l^I(x)}= \frac{\dd}{\dd \phi_l^I(x)}|_{j^I}+\frac{\p j^K(x)}{\p \phi_l^I(x)}\frac{\dd}{\dd j^K(x)}=\frac{\dd}{\dd \phi_l^I(x)}|_{j^I}+2\chi(x) \frac{\dd}{\dd j^I(x)}
\]

Let us evaluate the functional derivatives:

\[
\frac{\dd e^{-S_{\lm,I}[\phi_l,J]} }{\dd \phi_l^I(x)}= -\int {\cal D\chi}~[2 \chi(x)\phi_l^I(x)+2\chi(x)\int _y \big[ \frac{1}{\frac{\sqrt N}{\DD_h}+ 2\chi}\big]_{xy}j^I(y)]e^{-S_{\lm}[\chi,\phi_l,J]}
\]

\[
\frac{\dd^2 e^{-S_{\lm,I}[\phi_l,J]} }{\dd \phi_l^I(x)\dd \phi_l^I(y)}|_{\phi_l=0}=\int {\cal D\chi}~2N[2 \chi(x) \dd(x-y)+4 \chi(x) \big[ \frac{1}{\frac{\sqrt N}{\DD_h}+ 2\chi}\big]_{xy}\chi(y)]e^{-S_{\lm}[\chi,\phi_l,J]}
\]
So the ERG equation becomes:
\[
\frac{\p }{\p t}e^{-S_{\lm,I}}\Big|_{\phi_l=0}=\hf \frac{1}{\sqrt N}\int_x\int_y \dot \DD _{hxy}\frac{\dd^2}{\dd \phi_l ^I(x)\dd \phi_l ^I(y)} e^{-S_{\lm,I}}\Big|_{\phi_l=0}
\]
\[=
\hf \frac{1}{\sqrt N}\int_x\int_y \dot \DD _{hxy}\int {\cal D\chi}~2N[2 \chi(x) \dd(x-y)+4 \chi(x) \big[ \frac{1}{\frac{\sqrt N}{\DD_h}+ 2\chi}\big]_{xy}\chi(y)]e^{-S_{\lm}[\chi,0,J]}
\]
\be   \label{erg4}
=
\int {\cal D\chi}~\int _x \sqrt N[2 \chi(x) \dot\DD_h(0)+\int_x\int_y~4 \chi(x) \dot \DD _{hxy} \big[ \frac{1}{\frac{\sqrt N}{\DD_h}+ 2\chi}\big]_{xy}\chi(y)]e^{-S_{\lm}[\chi,0,J]}
\ee
 
 Let us expand 
 \[
 \big[ \frac{1}{\frac{\sqrt N}{\DD_h}+ 2\chi}\big]_{xy}=\frac{1}{\sqrt N}\big[ \frac{1}{\DD_h^{-1} + \frac{2 \chi}{\sqrt N}}\big]_{xy}
 \]
\[
=\frac{1}{\sqrt N}\big[\DD_{hxy}- \int_u~\DD_{hxu} \frac{2 \chi(u)}{\sqrt N}\DD_{huy} +     \int_u\int_v \DD_{hxu}\frac{2 \chi(u)}{\sqrt N}\DD_{huv}\frac{2 \chi(v)}{\sqrt N}\DD_{hvy}+...\big]
\]
Thus we get
\[
\frac{\p }{\p t}e^{-S_{\lm,I}[0,J]}=\int {\cal D}\chi~\{\int _x \sqrt N 2 \chi(x) \dot\DD_h(0)+
4 \int_x \int _y~\dot \DD_{hxy}~~\chi(x)\big[\DD_{hxy}- \int_u~\DD_{hxu} \frac{2 \chi(u)}{\sqrt N}\DD_{huy} 
\]
\be   \label{erg6}
+     \int_u\int_v \DD_{hxu}\frac{2 \chi(u)}{\sqrt N}\DD_{huv}\frac{2 \chi(v)}{\sqrt N}\DD_{hvy}+...\big]\chi(y)\}e^{-S_{\lm,I}[\chi,0,J]}
\ee

As mentioned above, $S_{\lm,I}[0,J]$ is the generating functional for the $\chi$ field theory and thus we can set $S_{\lm,I}[0,J]\equiv- W_\lm[J]$. Thus 
\[
\frac{\dd Z}{\dd J(x)}= \frac{\dd e^{W_\lm[J]}}{\dd J(x)}=<\chi(x)>|_J
\]...etc.

The first term on the right is $\sqrt N\dot \DD_h(0)\int_x\langle \chi(x) \rangle_J$. For the critical theory $\langle \chi(x) \rangle=0$ when $J=0$. More generally the first term is given by:
\[
2 \sqrt N \dot \DD_h(0)\int _x \frac{\dd e^{W_\lm[J]}}{\dd J(x)}
\]
The leading contribution is (see \eqref{Zleading} below)
\[
 \sqrt N \dot \DD_h(0) \int _x \int _y [\frac{1}{\DD_h^2}(x-y)]J(y)
\]
In momentum space this corresponds to
\[
\int _p \frac{1}{\DD_h^2(p)} J(p)\dd(p)=\frac{1}{\DD_h^2(0)} J(0)
\]
We will set
 \be  \label{const}
 J(0)=0
 \ee 
 and this term vanishes. We will assume that the external source satisfies this constraint of vanishing zero component of momentum. This means of course that we are restricted to calculations of correlations at non zero momenta, which is all we need.

The second term on the RHS of \eqref{erg6} that is quadratic in $\chi$ can thus be written  as:
\[
4 \int_x \int _y~\dot \DD_{hxy}~\int {\cal D}\chi~\chi(x)\DD_{hxy}\chi(y)e^{-S_{\lm,I}[\chi,0,J]}=4 \int_x \int _y~\dot \DD_{hxy}\DD_{hxy}\frac{\dd^2e^{W_\lm[J]}}{\dd J(x)\dd J(y)}
\]
and thus the ERG equation for $W_\lm[J]$  starts off as a  Polchinski's ERG equation as if  $\chi$ were a fundamental field sourced by $J$. 
\be
\frac{\p }{\p t}e^{W_\lm[J]}= 
2\int_x\int_y \dot {(\DD_{hxy})^2}\frac{\dd^2e^{W_\lm[J]}}{\dd J(x)\dd J(y)} +...
\ee

%

The higher order (in $\frac{1}{N}$ and $\chi$) terms modify this. The next term of $O(1/\sqrt N)$ is
\be    \label{erg8}
-8\frac{1}{\sqrt N} \int_x \int _y\int_u~\dot \DD_{hxy}\DD_{hxu} \DD_{huy}~\int {\cal D}\chi~\chi(x) \chi(u) \chi(y)e^{-S_{\lm,I}[\chi,0,J]}
\ee
Note that it is symmetric in $x,y,u$ and corresponds to the diagram in Figure 3.  It is useful to observe that because of the symmetry,
\be
\dot \DD_{hxy}\DD_{hxu} \DD_{huy}=-\frac{1}{3}\lm \frac{d}{d\lm}  (\DD_{hxy}\DD_{hxu} \DD_{huy})
\ee
This will be useful later on. In the ERG this higher order cubic term contributes a term proportional to three derivatives:

\[
\int {\cal D}\chi~\chi(x) \chi(u) \chi(y)e^{-S_{\lm,I}[\chi,0,J]}=\langle \chi(x) \chi(u) \chi(y)\rangle_J=\frac{\dd^3}{\dd J(x)\dd J(u)\dd J(y)}Z[J]
\]
Thus \eqref{erg8} becomes
\be   
-8\frac{1}{\sqrt N} \int_x \int _y\int_u~\dot \DD_{hxy}\DD_{hxu} \DD_{huy}~\frac{\dd^3}{\dd J(x)\dd J(u)\dd J(y)}Z[J]
\ee

and so \eqref{erg6} becomes, keeping terms to $O(1/\sqrt N)$:
\begin{tcolorbox}
\be   \label{erg8.5}
\frac{\p }{\p t}e^{W_{\lm}[J]}=\{4 \int_x \int _y~\dot \DD_{hxy}\DD_{hxy}\frac{\dd^2}{\dd J(x)\dd J(y)}-8\frac{1}{\sqrt N} \int_x \int _y\int_u~\dot \DD_{hxy}\DD_{hxu} \DD_{huy}~\frac{\dd^3}{\dd J(x)\dd J(u)\dd J(y)}+...\}e^{W_\lm[J]}
\ee
\end{tcolorbox}
At higher orders there will be higher derivatives. This equation is one of the main results in this paper and enables us to make contact with Holographic RG.

The triple derivative  and higher order terms are not there in the original Polchinski ERG equation. We will see below that these terms are crucial for making contact with Holographic RG encountered in the AdS/CFT correspondence. We can understand these terms as follows.  As outlined in Section \eqref{rev} the Wilson action for a fundamental scalar field is obtained by integrating out $\phi_h$. The kinteic term for $\phi_h$ is $\hf\phi_h \DD_h^{-1}\phi_h$ and depends on $\lm$ through $\DD_h$. The $\lm$ dependence of $S_{\lm,I}$ can all be traced to this term. On the other hand, the $\lm$ dependence of the action for $\chi$ involves all the interaction terms, in addition to the kinetic term $\chi \DD_h ^2 \chi$. This is because these terms were generated in the first place by integrating out $\phi_h$. Thus as $\chi$ is integrated out there are several sources of $\lm$ dependence in the Wilson action. The first term in the ERG equation \eqref{erg8.5} captures the $\lm$ deendence from the kinetic term, and all the higher terms capture the various other sources of $\lm$ dependence.

\subsection{Simplifying the Equation  in $1/N$ Expansion}

We would like a second order (functional) differential equation and  using the leading approximation in $O(1/N)$ to $Z[J]$ one can replace the higher derivative terms in \eqref{erg8.5} by polynomials in $J$.
We start with
\[
S_{\lm,I}[\chi,0,J]=NTr \ln (\frac{\sqrt N}{\DD_h}+ 2\chi)-\int \chi(J+\frac{\sqrt N}{2\bar u}r) +\frac{\chi^2}{2 \bar u}
\]
Expand in powers of $\chi$ to get (neglecting a field independent term):
\be   \label{erg9}
S_{\lm,I}[\chi,0,J]=2\sqrt N\int _x \DD_h(0)\chi - \frac{\sqrt N}{2\bar u}r\int _x \chi(x) - 2\int_x\int_y \chi(x)\DD^2_{hxy}\chi(y) + O(\chi^3/\sqrt N)-\int_x~J\chi
\ee

We choose $r=2\bar u\DD_h(0)=r_c$  so that the linear term in $S[\chi,0,0]$ vanishes. We have also taken $\bar u \to \infty$ which is the WF fixed point in 3 dimensions.

Then \footnote{Note that $\chi$ contour being along the imaginary axis gives the right sign for the kinetic term.}
\[
Z[J]= \int {\cal D}\chi e^{ 2\int_x\int_y \chi(x)\DD^2_{hxy}\chi(y) + O(\frac{1}{\sqrt N}\chi^3)+\int_x~J\chi}
\]
\be    \label{Zleading}
= e^{ -\int_x\int_y J(x)\big[\frac{1}{8 \DD_h^2}\big]_{xy}J(y) + O(\frac{1}{\sqrt N})}
\ee
is the leading approximation to $Z[J]$.

We can also check that \eqref{Zleading} solves the ERG equation to this order. The leading term in the ERG equation is of the Polchinski form:
\be    \label{erg8.55}
\frac{\p }{\p t}\psi[J,t]=
2\int_x\int_y \dot {{\cal G}(t)^2}\frac{\dd^2\psi[J,t]}{\dd J(x)\dd J(y)}
\ee
The evolution operator is
\be \label{evop2}
\psi[J_f,t_f]= \int {\cal D}J_ie^{-\frac{1}{8}\frac{(J_f-J_i)^2}{{\cal G}(t_f)-{\cal G}(t_i)}}\psi[J_i,t_i]
\ee
It is easy to check that 
\be \label{erg8.6}
\psi[J,t]= e^{-\frac{1}{8}\frac{J^2}{{\cal G}(t)}}
\ee
solves this. This is exactly as given by \eqref{Zleading}. 

Now we use the leading order solution to estimate the higher derivative correction term of $O(1/\sqrt N)$. Acting on the leading $O(1)$ solution \eqref{Zleading} one obtains:
\[
\frac{\dd^3}{\dd J(x)\dd J(u)\dd J(y)}Z[J]=-\big[\frac{1}{4 \DD_h^2}(J)\big]_x\big[\frac{1}{4 \DD_h^2}(J)\big]_u\big[\frac{1}{4 \DD_h^2}(J)\big]_y -\big[\frac{1}{4 \DD_h^2}\big]_{xu}\big[\frac{1}{4 \DD_h^2}(J)\big]_y +O(\frac{1}{\sqrt N})Z[J]
\]

Thus the higher derivative term in \eqref{erg8.5} becomes

\be   \label{hd}
 \frac{1}{\sqrt N}\int_x \int _y\int_u~\dot \DD_{hxy}\DD_{hxu} \DD_{huy}\big[\frac{1}{2 \DD_h^2}(J)\big]_x\big[\frac{1}{2 \DD_h^2}(J)\big]_u\big[\frac{1}{2 \DD_h^2}(J)\big]_y +2\big[\frac{1}{2 \DD_h^2}\big]_{xu}\big[\frac{1}{2 \DD_h^2}(J)\big]_y Z[J]+O(\frac{1}{ N})
\ee

This corresponds to Figure 3 with $J$'s attached at each end and a propagator connecting it. The linear term in $J$ corresponds to a closed loop and
 the graph is shown in Figure 2. It corresponds to a zero momentum $J$ and vanishes by constraint \eqref{const}.
 
\begin{figure}
\includegraphics[width=18cm]{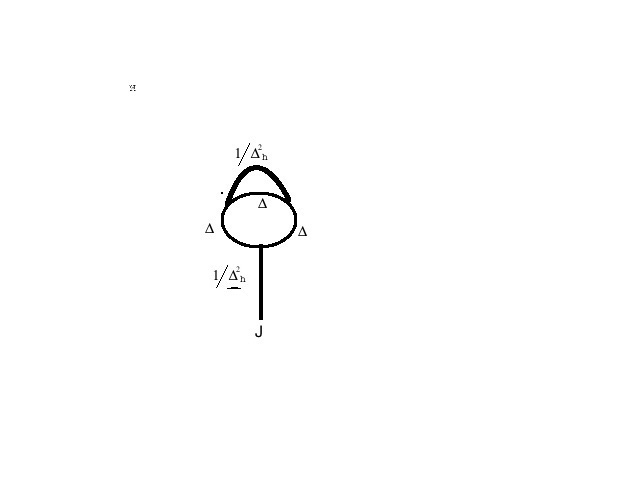}
\caption{The linear term in J vanishes because $J(0)=0$}
\end{figure}

Thus substituting \eqref{hd} minus the linear term, in \eqref{erg8.5} we get
\[
\frac{\p }{\p t}e^{W_\lm[J]}=
2\int_x\int_y \dot {(\DD_{hxy}^2)}\frac{\dd^2e^{W_\lm[J]}}{\dd J(x)\dd J(y)} 
+8 \frac{1}{\sqrt N}\int_x \int _y\int_u~\dot \DD_{hxy}\DD_{hxu} \DD_{huy}~\int {\cal D}\chi~\chi(x) \chi(u) \chi(y)e^{-S_{\lm,I}[\chi,0,J]}
\]
\[
=
\{2\int_x\int_y \dot {(\DD_{hxy}^2)}\frac{\dd^2}{\dd J(x)\dd J(y)} + \frac{1}{\sqrt N}\int_x \int _y\int_u~\dot \DD_{hxy}\DD_{hxu} \DD_{huy}\big(\big[\frac{1}{2 \DD_h^2}(J)\big]_x\big[\frac{1}{2 \DD_h^2}(J)\big]_u\big[\frac{1}{2 \DD_h^2}(J)\big]_y 
\]

\be   \label{erg10}
\ee

\eqref{erg10} is a generalization to auxiliary fields such as $\chi$, of Polchinski ERG equation for elementary fields and is a principal result of this paper.
Given an action as in \eqref{erg9} with a kinetic term $\DD_h^2$, the Polchinski ERG equation for a situation where $\chi$ is a fundamental field consists of only the first term in
\eqref{erg10}. However since $\chi$ is actually stands for a composite operator $\phi^2$, all the terms in the action $S_\lm[\chi,0,J]$ for $\chi$ contain a $\lm$ dependence. Thus the ERG equation \eqref{erg10} includes the potential term  $O(J^3)$ and higher.  Interestingly, this is precisely the kind of term one expects in Holographic RG equations that show up in AdS/CFT calculations. 

To this order in $1/N$ our ERG equation for $W_\lm[J]$ is thus of the form:
\bc
\be \label{erg11}
\frac{\p }{\p t}e^{W_\lm[J]}=
\{2\int_x\int_y \dot {(\DD_{hxy}^2)}\frac{\dd^2}{\dd J(x)\dd J(y)}+ \frac{1}{\sqrt N}\int_x\int_y\int_z f(x,y,z,\lm) J(x)J(y)J(z)\}e^{W_\lm[J]}
\ee
\ec

with
\be  \label{f}
f(x,y,z,\lm)=  \int_{x'} \int _{y'}\int_{z'}~\dot \DD_{hx'y'}\DD_{hx'z'} \DD_{hz'y'}\big[\frac{1}{2 \DD_h^2}\big]_{x'x}\big[\frac{1}{2 \DD_h^2}\big]_{y'y}\big[\frac{1}{2 \DD_h^2}\big]_{z'z}
\ee
\begin{figure}
\hspace{5cm}\includegraphics[width=15cm]{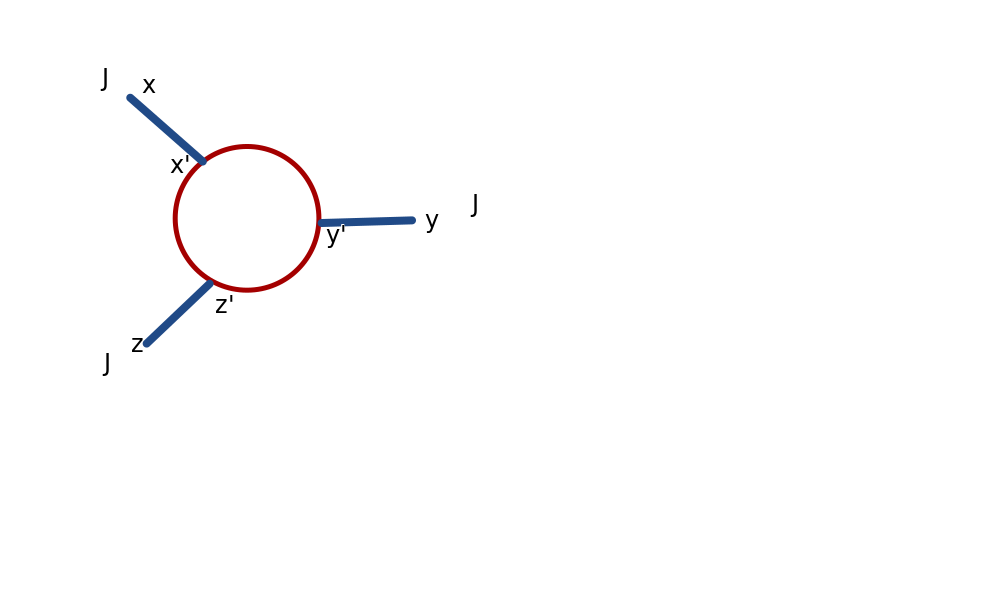}\hspace*{\fill}
\caption{Cubic Term. The external legs are $\chi$ propagators $\frac{1}{\DD_h^2}$ and the internal lines are $\phi$ propagators $\DD_h$.}
\end{figure}

The product of three $\DD$'s corresponds to the loop in Figure 3. Note that $\DD_h$ is the high energy propagator  for $\phi_h$ - see \eqref{erg1}. The objects in square brackets are the  $\chi$ propagators - see \eqref{erg9} - on the external legs. The $\chi$ propagator $\frac{1}{\DD_h^2}$ is generated by the  diagram in Figure 4.
\begin{figure}
\hspace{4cm}\includegraphics[width=10cm]{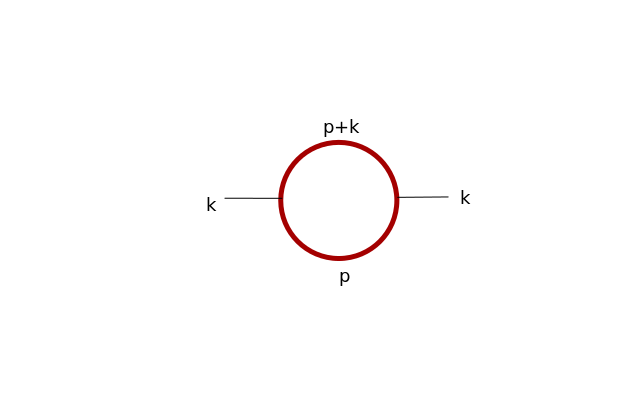}
\caption{Kinetic term for $\chi$. The internal lines correspond to $\phi$ propagators $\DD_h$.}
\end{figure}

For the ERG \eqref{erg11} the precise cutoff dependence is important. Thus  for instance a standard choice for $\DD_h$ is 
\[
\DD_h(p) = \frac{K_0(p)-K(p)}{p^2}
\]
with $K(p) = e^{-p^2/\lm^2}$ and $K_0(p) = e^{-p^2/\lo^2}$. With this choice the loop integrals do not give  convenient analytical expressions. Note however, that $\lo\to \infty$ is the continuum limit and $\lm \to 0$ is the limit where the full functional integral is done, and in this limit the loop integrals can be done. 

 If the theory is {\em exactly} at a fixed point we expect the cutoff dependence to be determined by the scaling dimensions. {\em Near} the fixed point one can expect logarithmic deviations. These logarithms are universal and do not depend on the details of the cutoff scheme. Keeping this in mind we will evaluate these loops with some convenient cutoff scheme. Appendix \eqref{A} contains some details of these calculations.

 The result  in $D=4-2\eps$ for $S_2[\chi]$ which defines $\DD_h^2$ is
 \be \label{dh1}
S_2[  \chi ]= \frac{1}{(4\pi)^{\frac{D}{2}}}\frac{\Gamma (\Dt+1)\Gamma(\hf)}{\Gamma(\frac{D+3}{2})}\int _k~  \chi(k) [2(k\lo)^{-\eps} K_\eps(\frac{2k}{\lo}) -2(k\lm)^{-\eps} K_\eps(\frac{2k}{\lm}) ]  \chi(-k)+\frac{1}{2\bar u} \int _k~ \chi(k)  \chi(-k) 
\ee

If we take $\lo\to \infty$ and $\lm\to 0$ one obtains
\[
S_2[  \chi]= \frac{1}{(4\pi)^{\frac{D}{2}}}\frac{\Gamma (\Dt+1)\Gamma(\hf)}{\Gamma(\frac{D+3}{2})} 2^{-1-\eps}\Gamma(\eps)\int _k   \chi(k)[   k^{-2\eps} ]  \chi(-k)
\]
\[
= \frac{1}{2\gamma(D)}\int \frac{d^3k}{(2\pi)^3}   \chi(k) k^{-1}   \chi(-k)
\]
so that
\[
\gamma(D)=\frac{\Gamma(\frac{D+3}{2})}{\Gamma (\Dt+1)\Gamma(\hf)}2^\eps (4\pi)^{\frac{D}{2}}
\]
Finally  if $D=3$ so that $\eps=\hf$ one obtains:
\be    \label{s2}
S_2[\chi]=\frac{3}{128\sqrt\pi}\int \frac{d^3k}{(2\pi)^3} \chi(k) k^{-1}\chi(-k)
\ee

Thus 
\be   \label{dh2}
\lim _{\lm\to 0,\lo\to \infty}\DD_h^2(k)=\gamma(3) k^{-1}
\ee
Also from \eqref{s2} we see that $[\chi(k)]=-1$ which gives $[\chi(x)]=2$ as expected.

In the cubic term we need to evaluate the loop diagram in Figure 3, which is done in Appendix \eqref{B}. Once again we are interested in the near fixed point behaviour so we use a cutoff scheme that is amenable to analytic evaluation. What we need is (assuming symmetry in $x',y',z'$):
\[
\dot \DD_{hx'y'}\DD_{hx'z'} \DD_{hz'y'} =\frac{1}{3}\frac{d}{dt}(\DD_{hx'y'}\DD_{hx'z'} \DD_{hz'y'})
\]

The loop diagram gives rise to the following term in the Wilson action for $\chi$:
\[
S_3[\chi]=\frac{1}{3!\sqrt N}\int_x\int_y\int_z ~\DD_{hx y}\DD_{hxz} \DD_{hzy}\chi(x)\chi(y)\chi(z)
\]
 Now as mentioned above, $\DD_{hx'y'}\DD_{hx'z'} \DD_{hz'y'}$ for general IR cutoff $\lm$ is complicated. It turns out (see below) that we will actually need only the value at $\lm=0$.
The result, for energies much less than the UV cutoff $\lo$, can be summarized by the following scale invariant term in the action for $\chi$, in momentum space:
\[
S_3[\chi]=\frac{1}{\sqrt N}\frac{1}{3!}\int _{k_1}\int_{k_2}\int_{k_3}\frac{1}{8k_1k_2k_3}\chi(k_1)\chi(k_2)\chi(k_3)\dd(k_1+k_2+k_3)
\]

Note that the $\chi$ propagator actually diverges at high momenta, but the interactions go to zero and so the theory is well behaved at high energies. One can do a field redefinition
to make this manifest but we will not bother about this here.
\section{Holographic Rewriting}
In \cite{Sathiapalan:2017} the Polchinski ERG evolution operator of a D-dimensional field theory was written as a D+1 dimensional
field theory (“bulk”) functional integral. This is a holgraphic version of ERG and is very similar to holographic
RG in AdS/CFT correspondence. In this section we do the same for the ERG equation \eqref{erg11}. The new
ingredient is the presence of potential terms in the D+1 dimensional bulk action. This makes the similarity
with holographic RG stronger. We write down the bulk action and also evaluate two and three point correlation
functions using a semiclassical solution of the bulk theory. This is valid in the large N limit. Standard results
of the $O(N )$ model are reproduced. This is of course built into the construction since it is just a rewriting of
the usual ERG equations.

The bulk action however has a non standard kinetic term. In the next section by a field redefinition this
will be converted to a standard kinetic term - but in AdS space.

\subsection{Holographic Action}
The ERG equation \eqref{erg11} for which an evolution operator is required is of the form
\be   \label{erg13}
\frac{d\psi}{dt}= (2 \dot {\cal G} \frac{\p^2}{\p x^2} + \frac{1}{\sqrt N}V(x))\psi
\ee
where ${\cal G}=\DD^2_h$.
And the evolution operator is of the form, by analogy with the Feynman Path Integral solution for the Schroedinger equation:
\[
\psi(x_f,T)=\int dx_i\int_{x(0)=x_i,~x(T)=x_f} {\cal D}x(t)e^{\int _0^Tdt~[ -\frac{\dot x^2}{8\dot {\cal G}}+\frac{1}{\sqrt N}V(x(t))]}\psi(x_i,0)
\]
\[=
\int dx_i~ U(x_f,T;x_i,0)\psi(x_i,0)
\]
Thus as mentioned above, the evolution operator is a $D+1$ dimensional field theory (in our case $D=3$) and can be thought of as  a``holographic RG" where the radial direction is $t=\ln \lo/\lm$.
The expression for the evolution operator to this order in $1/N$ is:
\bc
\be  \label{evop}
U[J_f,t_f;J_i,t_i]=\int {\cal D} J(k,t)e^{ \int dt~\big[ \int _k -\frac{\dot J(k,t)\dot J(-k,t)}{8\dot \DD_h^2(k,t)} + \frac{1}{\sqrt N}\int_{k_1,k_2,k_3}\dd(k_1+k_2+k_3)f(k_1,k_2,k_3,t)J(k_1)J(k_2)J(k_3)+...O(1/N)\big]}
\ee
\ec
where $f$ is given in position space in \eqref{f}. $\DD^2_h$ is determined by the one loop diagram in Figure 3. Approximate expressions are given in \eqref{dh1},\eqref{dh2}.

We would like to now point out an important difference with the discussions in  \cite{Sathiapalan:2017,Sathiapalan:2019}. In
\cite{Sathiapalan:2017} the  holographic field theory had only a kinetic term - the counterpart of $\hf \frac{\dot x^2}{\dot {\cal G}}$ here. Thus the only information it had about the boundary field theory was the function ${\cal G}$.  ${\cal G}$ is related to the propagator of the free field theory for $\chi$. One of the open problems discussed in \cite{Sathiapalan:2017,Sathiapalan:2019} was precisely this: the bulk action was universal, in that it had no dependence on the interaction terms of the boundary field theory. This reflects the fact that the Wilsonian ERG evolution operator is universal - it is the same for all field theories. It only depends on the kinetic term. On the other hand in holographic RG as described in the AdS/CFT correspondence, the bulk action has {\em all the information} about the boundary theory and is thus a ``dual" theory. This problem was discussed in \cite{Sathiapalan:2017,Sathiapalan:2019} and left as an open problem, although a (different) solution was also proposed.

In the the ERG equation considered in this paper \eqref{erg10},  and the corresponding bulk action given in \eqref{erg13} there are potential terms.  {\em Thus all the information of the boundary theory is present in a natural way} and it is thus a candidate for being a true holographic dual of the boundary theory. The main difference with \cite{Sathiapalan:2017} is that the boundary operators being discussed are composites. This is quite natural in the context of AdS/CFT. 

 The auxiliary field $\chi$ corresponds to the  operator $\phi^2$. So our holographic theory has information only about correlators of this operator. The original $O(N)$ model has an infinite number of such operators. Corresponding to each operator one can introduce an auxiliary field and write a holographic bulk field theory describing the ERG evolution. Thus the bulk theory would have an infinite number of fields. This is all exactly the same in the AdS/CFT context also. In particular there is a spin 2 auxiliary field (metric field) corresponding to the composite operator $T_{\mu\nu}$ - the energy momentum tensor. The bulk theory describing this is expected to be the Einstein action.

\subsection{ Solution in $1/N$ using Semiclassical Approximation}

We would like to solve 
\be
\frac{\p }{\p t}e^{W_\lm[J]}=
\{2\int_x\int_y \dot {(\DD_{hxy}^2)}\frac{\dd^2}{\dd J(x)\dd J(y)}+ \frac{1}{\sqrt N}\int_x\int_y\int_z f(x,y,z,\lm) J(x)J(y)J(z)\}e^{W_\lm[J]}
\ee
The Functional Integral representation \eqref{evop} reproduced below. We evaluate semiclassically by solving the EOM and plugging back into the action defined below.
\br   \label{evop1}
U[J_f,t_f;J_i,t_i]&=&\int {\cal D} J(k,t)e^{ \int dt~\big[ \int _k -\frac{\dot J(k,t)\dot J(-k,t)}{8\dot {\cal G}(k,t)} + \frac{1}{\sqrt N}\int_{k_1,k_2,k_3}\dd(k_1+k_2+k_3)f(k_1,k_2,k_3,t)J(k_1)J(k_2)J(k_3)+...O(1/N)\big]}\nonumber \\
&\equiv& \int {\cal D} J(k,t)e^{-S[J]}
\er
In the semiclassical approximation
\be
U[J_f,t_f;J_i,t_i]=e^{-S_{cl}[J_f,t_f;J_1,t_i]}
\ee
\subsubsection{$O(1)$}
The $O(1)$ term is quadratic and here the answer is exact. Note that we have suppressed the momentum label, $p$ (and the integration over $p$)  of $J(p)$ for convenience. They will be restored later.

\be  \label{evop3.1}
-\int_{t_i}^{t_f} dt  \frac{\dot J^2}{8\dot{\cal G}}
\ee

Take $\frac{1}{\lm} =z=\mu e^t$. $\mu$ is an arbitrary reference scale that can even be set to 1.  So $z=0$ is $t=-\infty$ and $z=\infty$ is $t=\infty$. We also let $\mu e^{t_i}=\eps$ and $\mu e^{t_f}=z_0$. The IR end can be taken as $z=z_0$. We would like to take $\eps \to 0$ ($\lo\to \infty$) and $z_0\to \infty$  at the end.
The EOM is
\be  \label{erg13.1}
\frac{d}{dt}\frac{\dot J}{\dot {\cal G}}=0 \implies \frac{\dot J}{\dot {\cal G}}=b=~const~\implies J=b{\cal G}+ c
\ee
where $b,c$ are some constants. 
\be    \label{erg13.2}
b= \frac{J(t_f)-J(t_i)}{{\cal G}(t_f)-{\cal G}(t_i)}
\ee
What about the boundary condition?

%
%
%

 One can posit a natural choice for the boundary functional for $\eps\to 0$ 
\be \label{bfnl}
\lim_{\eps\to 0}W_\eps[J]=\hf \frac{J(\eps)^2}{{\cal G}(\eps)}
\ee
Since $\lim_{\eps\to 0}{\cal G}(\eps)=0$ this becomes a delta function at the boundary.
In the semiclassical approximation when we solve the EOM 
we also get a boundary condition at $\eps\to 0$:
\be
-\dd J(\eps) \frac{\dot J(\eps)}{\dot {\cal G}(\eps)} +\dd J(\eps) \frac{J(\eps)}{{\cal G}(\eps)}=0
\ee
This equation implies
\be  \label{b}
b=\frac{J(\eps)}{{\cal G}(\eps)} 
\ee
Note that the boundary functional \eqref{bfnl} vanishes  at $\eps=0$. This is correct
since no integration of field modes has been done at $z=0$ (which is $\lm=\infty$).

Substituting in the action $S[J]$ we get for the free theory the classical expression
\be
S_{cl}[J_f,t_f;J_i,t_i]= \frac{1}{8} \int_{t_i}^{t_f}\frac{J\dot J}{8\dot {\cal G}}= \frac{1}{8}b (J(t_f)-J(t_i))= \frac{1}{8}\frac{(J(t_f)-J(t_i))^2}{{\cal G}(t_f)-{\cal G}(t_i)}
\ee
reproducing the earlier result \eqref{evop2}.
Let us take  $t_i=-\infty$ (i.e. $\lm =\infty,~z=0$) $t_f=\infty$ . Then 
\be
S_{cl}=\frac{1}{8}bJ(\infty)=\frac{1}{8}\frac{J(\infty)^2}{{\cal G}(\infty)}
\ee
Calling $J(\infty)=J_c$
\be
S_{cl}[J_c]=\frac{1}{8}b J_c= \frac{1}{8} J_c^2p
\ee
in $D=3$, 
which is as expected.

Thus in the leading semiclassical approximation (restoring the momentum labels):
\be   \label{erg18}
W[J]= \lim _{\lm\to 0} W_\lm[J]= -\frac{1}{8} \int _p J(p)J(-p)p
\ee

\subsubsection{Rewriting $O(1)$}

In the last section we saw that the final answer for $\psi(J)$:
\[
\psi(J_f,z_f)=\int dJ_i~ U(J_f,z_f,J_i,z_i)\psi(J_i,z_i)
\]
is a functional of $J_f=J(z_f)$. And the classical EOM was solved with $J(0)=0, ~J(\infty)=J_f=J_c$. This is normal in the functional formalism. However in AdS/CFT holography,in the ``standard quantization" the final answer is a functional of $J(z_i)$, where $z_i=\eps$ is at or very close to the boundary, because the bulk gravity action is evaluated with a specified value for all quantitites at the boundary  of AdS space. Accordingly we will rewrite our solution above using a modified form of the equation that does not alter the physical content.

Let us define, in $D=3$ \footnote{When $D=3+\delta$,  $\frac{1}{p}$ is replaced by $\frac{1}{p^{1-\dd}}$} 
\be  \label{G}
G(z)=\frac{\gamma(3)}{p}-{\cal G}(p,z)=\frac{\gamma(3)}{p}-\DD_h^2(p,z)
\ee
Since $\cal G$ is the contribution from modes integrated out between $\lm$ and $\infty$, what remains is the contribution from $0$ to $\lm$. So $G$ is the kinetic term generated for $\chi$ due to the $\phi$ modes below $\lm$ and $\lm$ thus acts as a UV cutoff for $G$.

It satisfies $G(\infty)=0$ (i.e. at $\lm=0$) and $G(0)=\frac{\gamma(3)}{p}$ ( corresponding to $\lm=\infty$).
Note also that 
\[
\dot{\cal G} = -\dot G
\]
%
%

Furthermore in the ERG equation the leading $O(1)$ term in \eqref{erg8.5} can be rewritten using $G$ (in momentum space)
\be  \label{erg19}
\frac{\p }{\p t}e^{W_{\lm}[J]}=-2 \int_p~\dot G(p,z)\frac{\dd^2}{\dd J(p)\dd J(-p)}e^{W_\lm[J]}
\ee

We take a solution to this  of the  form in \eqref{bfnl}
\be
\psi[J,t]= e^{\frac{1}{8}\frac{(J)^2}{ G(z)}}
\ee
This ensures that $J(\infty)=0$ as required.

The leading order action functional can be written in a form similar to \eqref{evop} except for replacing $\cal G$ by $G$:
\be \label{evop3}
\int _{t_i}^{t_f}\frac{1}{8}\frac{\dot J^2}{\dot G}
\ee
The solutions are the same $\dot J = b \dot G$ and so
\be
J= bG +c
\ee
Using $G(\infty)=J(\infty)=0$ we get $c=0$. Thus we get the same solutions parametrized by the value of $J(0)=J_c$.
\be
J(z)=bG(z)=J_c \frac{p}{\gamma(3)} G(z)
\ee
The value of the action in the semiclassical approximation is
\be
\frac{J\dot J}{8\dot G}\big|_{z=0}^{z=\infty}= -\frac{J_c(p)J_c(-p)p}{8\gamma(3)}
\ee 
as before. But the main difference with the calculation the last section is that the contribution comes from $z=0$ rather than $z=\infty$ - unlike the earlier calculation. This is just a different choice of variables. The only advantage of this choice is that it is similar to what happens in AdS/CFT calculations.
 
\subsubsection{$O(1/\sqrt N)$}
We now proceed to include the cubic term which is $O(\frac{1}{\sqrt N})$. We use the symbol $g$ for $\frac{1}{\sqrt N}$ for simplicity.

Write 
\[
J=J_0+ g J_1
\]
Keeping terms of $O(g)$ in the EOM obtained from \eqref{evop1} we get
\be  \label{erg20}
\frac{1}{4}\p_t (\frac{\dot J_1(p,t)}{\dot {\cal G}(p)}) + 3g \int_{k_1}\int_{k_2}f(p,k_1,k_2) J_0(k_2,t)J_0(p-k_2,t)\dd(p+k_1+k_2)=0
\ee
We further assume that $J_0(0)=0$ is the boundary condition at $z=0$ ($t=-\infty$) as seen from the last subsection. We will also assume that $J(\infty)=J_0(\infty)$ so that $J_1(\infty)=0$. We will further impose $\dot J_1(\infty)=0$ as a convenient choice because it removes a boundary term from the action (see below).

We rewrite the $O(g)$ term in the kinetic term of the action $S[J]$ as
\be  \label{erg22}
S_1=\int dt~ \frac{1}{4}\frac{\dot J_0\dot J_1}{\dot{\cal G}} = -\frac{1}{4}\int dt~J_0  \p_t (\frac{\dot J_1}{\dot{\cal G}})
+\frac{1}{4} \frac{J_0\dot J_1}{\dot {\cal G}}\Big|_{t=-\infty}^{t=\infty} 
\ee
The boundary conditions $\dot J_1(\infty)=0$ and $J_0(t=-\infty)=0$ get rid of the boundary term in \eqref{erg22}. Using \eqref{erg20} in \eqref{erg22} we get, putting the limits on the $t$ integration 
\be
S_1=-3g\int_{-\infty}^\infty dt ~ \int _p\int _{k_1} f(k_1,p-k_1,p,t)J_0(k_1,t)J_0(p-k_1,t)J_0(-p,t)
\ee

The interaction term in $S[J]$ contributes an $O(g)$ term
\be
S_2=g\int_{-\infty}^\infty dt~\int_{k_1,k_2,k_3}\dd(k_1+k_2+k_3)f(k_1,k_2,k_3,t)J_0(k_1)J_0(k_2)J_0(k_3)
\ee
The total is thus the $O(g)$ contribution to $S_{cl}$:
\[
S_{cl}=S_1+S_2=-2g\int_{-\infty}^\infty dt ~\int_{k_1,k_2,k_3}\dd(k_1+k_2+k_3)f(k_1,k_2,k_3,t)J_0(k_1,t)J_0(k_2,t)J_0(k_3,t)
\]
\be
\equiv -W^1[J_0]
\ee
$f$ is given by \eqref{f}, in position space:  
\be  
f(x,y,z,\lm)=  \int_{x'} \int _{y'}\int_{z'}~\dot \DD_{hx'y'}\DD_{hx'z'} \DD_{hz'y'}\big[\frac{1}{2 \DD_h^2}\big]_{x'x}\big[\frac{1}{2 \DD_h^2}\big]_{y'y}\big[\frac{1}{2 \DD_h^2}\big]_{z'z}
\ee
In momentum space it is 
\[
f(k_1,k_2,k_3,t)= \int _p \dot \DD_h(p)\DD_h(p+k_1)\DD_h(p+k_1+k_2) \frac{1}{{\cal G}(k_1,t){\cal G}(k_2,t){\cal G}(k_3,t)}  
\]
So $S_{cl}$ becomes
\[
S_{cl}=   2g\int_0^\infty dt~  \int_{k_1,k_2,k_3}\dd(k_1+k_2+k_3)\int _p \dot \DD_h(p)\DD_h(p+k_1)\DD_h(p+k_1+k_2) \frac{1}{{\cal G}(k_1,t){\cal G}(k_2,t){\cal G}(k_3,t)} J_0(k_1,t)J_0(k_2,t)J_0(k_3,t)
 \]  Using the symmetry of the loop diagram this can be written as:
 \be   \label{scl}
 S_{cl}=
\frac{2g}{3}\int_0^\infty dt~  \int_{k_1,k_2,k_3}\dd(k_1+k_2+k_3)\int _p \frac{d}{dt} (\DD_h(p)\DD_h(p+k_1)\DD_h(p+k_1+k_2)) \frac{1}{{\cal G}(k_1,t){\cal G}(k_2,t){\cal G}(k_3,t)} J_0(k_1,t)J_0(k_2,t)J_0(k_3,t) 
\ee
Now note that $\frac{J_0(k,t)}{{\cal G}(k,t)}=b(k)$ is independent of $t$. Thus at this order the integrand is a total derivative in $t$. Since $\DD_h$ vanishes at $z=0$ only the term at $z=\infty$ contributes. 
\be
S_{cl}=\frac{2g}{3}~  \int_{k_1,k_2,k_3}\dd(k_1+k_2+k_3)\int _p [ (\DD_h(p)\DD_h(p+k_1)\DD_h(p+k_1+k_2))]\Big|_{t=\infty} \underbrace{\frac{1}{{\cal G}(k_1,t){\cal G}(k_2,t){\cal G}(k_3,t)} J_0(k_1,t)J_0(k_2,t)J_0(k_3,t)}_{=b(k_1)b(k_2)b(k_3)~(t-independent)}
\ee
$b$ can also be evaluated at  $z=\infty$ (i.e. $t=\infty$) and we get using $J(\infty)=J_c$ :
\[
-W_1[J_c]=S_{cl}[J_c]
\]
where
\be
S_{cl}=\frac{2g}{3}~  \int_{k_1,k_2,k_3}\dd(k_1+k_2+k_3)\int _p  (\DD_h(p)\DD_h(p+k_1)\DD_h(p+k_1+k_2))\Big|_{t=\infty} \frac{1}{{\cal G}(k_1,\infty){\cal G}(k_2,\infty){\cal G}(k_3,\infty)} J_c(k_1)J_c(k_2)J_c(k_3)
\ee
And in position space, by the same logic:
\be   \label{erg24}
S_{cl}=\frac{2}{3\sqrt N} \int_{x',y',z'} \int _{x,y,z}~ [\DD_{hx'y'}\DD_{hx'z'} \DD_{hz'y'}]\Big|_{t=\infty}\big[\frac{1}{2 \DD_h^2}\big]_{x'x}\big[\frac{1}{2 \DD_h^2}\big]_{y'y}\big[\frac{1}{2 \DD_h^2}\big]_{z'z}J_c(x)J_c(y)J_c(z)
\ee

\subsubsection{Rewriting $O(1/\sqrt N)$ using $G$ instead of ${\cal G}$}

The calculation is more or less the same, except that contributions are picked up from $z=0$ rather than $z=\infty$.
Write, as before
\[
J=J_0+ g J_1
\]
Keeping terms of $O(g)$ in the EOM obtained from \eqref{evop1} we get
\be  \label{erg25}
\frac{1}{4}\p_t (\frac{\dot J_1(p,t)}{\dot  G(p)}) + 3g \int_{k_1}\int_{k_2}f(p,k_1,k_2) J_0(k_2,t)J_0(p-k_2,t)\dd(p+k_1+k_2)=0
\ee
We further assume that $J_0(0)=J_c$ is the boundary condition at $z=0$ so that $J_1(0)=0$. We will also assume that $J(\infty)=J_0(\infty)=0$. We will further impose $\dot J_1(0)=0$ as a convenient choice because it removes a boundary term from the action: 
\be  \label{erg25.5}
S_1=-\int dt~ \frac{1}{4}\frac{\dot J_0\dot J_1}{\dot G} =\frac{1}{4}\int dt~J_0  \p_t (\frac{\dot J_1}{\dot G})
-\frac{1}{4} \frac{J_0\dot J_1}{\dot  G}\Big|_0^\infty 
\ee
The boundary conditions $\dot J_1(0)=0$ and $J_0(\infty)=0$ get rid of the boundary term in \eqref{erg25.5}.

The rest of the calculations go through exactly as in the previous section with $G$ replacing $\cal G$.
We get in place of \eqref{scl}
 \be   
 S_{cl}=
\frac{2g}{3}\int_{-\infty}^\infty dt~  \int_{k_1,k_2,k_3}\dd(k_1+k_2+k_3)\int _p \frac{d}{dt} (\DD_h(p)\DD_h(p+k_1)\DD_h(p+k_1+k_2)) \frac{1}{ G(k_1,t)G(k_2,t)G(k_3,t)} J_0(k_1,t)J_0(k_2,t)J_0(k_3,t) 
\ee
Since$\frac{J}{G}=b$ is constant we get 
\be
S_{cl}=\frac{2g}{3}~  \int_{k_1,k_2,k_3}\dd(k_1+k_2+k_3)\int _p  (\DD_h(p)\DD_h(p+k_1)\DD_h(p+k_1+k_2))\Big|_{t=\infty} \underbrace{\frac{1}{G(k_1,t)G(k_2,t)G(k_3,t)} J_0(k_1,t)J_0(k_2,t)J_0(k_3,t)}_{=b(k_1)b(k_2)b(k_3)}
\ee
The value of $b$ can be evaluated at $z=0$ or $t=-\infty$, and is seen to be $\frac{J(0)}{G(0)}=J_cp$ (when $D=3$).

And in position space ($J_0(0)=J_c$):
\be   
S_{cl}=\frac{2}{3\sqrt N} \int_{x',y',z'} \int _{x,y,z}~ \DD_{hx'y'}\DD_{hx'z'} \DD_{hz'y'}\big[\frac{1}{2 \DD_h^2}\big]_{x'x}\big[\frac{1}{2 \DD_h^2}\big]_{y'y}\big[\frac{1}{2 \DD_h^2}\big]_{z'z}J_c(x)J_c(y)J_c(z)
\ee
exactly as in \eqref{erg24}.

The difference is that the final answer is parametrized by the value of $J$ at the boundary $z=0$ rather than at infinity.
This makes the comparison with AdS/CFT a little easier.

\subsubsection{Three Point Correlator in Position Space}

We need to evaluate the $O(g)$ cubic term in $W_\lm[J]$ with $\lm \to 0$.
The calculation will be done in $D$ dimensions. Since we have $\lm \to 0$ we get the usual propagators.

$\chi(x)$ has dimension 2 and $\phi(x)$ has dimension $\Dt-1$. So $\langle \phi(x) \phi(0)\rangle \approx \frac{1}{x^2(\Dt-1)}$ and $\langle \chi(x)\chi(0) \rangle \approx \frac{1}{x^4}$.

The integral to be done is
\be
I=\int d^Dy_1\int d^Dy_2\int d^Dy_3~ \frac{1}{(x_1-y_1)^4(x_2-y_2)^4(x_3-y_3)^4 (y_1-y_2)^{2(\Dt-1)}(y_3-y_2)^{2(\Dt-1)}(y_1-y_3)^{2(\Dt-1)}}
\ee
Use the result
\[
\int d^Dy \frac{1}{(x_1-y)^{2d_1}(x_2-y)^{2d_2}(x_3-y)^{2d_3}}=
\]
\be
 \frac{1}{(x_1-x_2)^{D-2d_3}(x_1-x_3)^{D-2d_2}(x_3-x_2)^{D-2d_1}}\frac{\Gamma(\Dt-d_1)\Gamma(\Dt-d_2)\Gamma(\Dt-d_3)}{\Gamma(d_1)\Gamma(d_2)\Gamma(d_3)}
\ee

and do the $y_1 , y_2 , y_3$ integrals one after the other to get
\be
I= \frac{1}{(x_1-x_2)^{2}(x_1-x_3)^2(x_2-x_3)^2}\frac{\Gamma(\Dt-2)^3\Gamma(3-\Dt)}{\Gamma(\Dt-1)^3 \Gamma(D-3)}
 \ee
  and 
  \be  \label{erg30}
  W_1[J_c]= \frac{2}{3\sqrt N}\frac{\Gamma(\Dt-2)^3\Gamma(3-\Dt)}{\Gamma(\Dt-1)^3 \Gamma(D-3)}
\int _{x_1}\int _{x_2}\int _{x_3}\frac{1}{(x_1-x_2)^{2}(x_1-x_3)^2(x_2-x_3)^2}J(x_1)J(x_2)J(x_3)
\ee
Note that in $D=3$ it vanishes.  This result is known  \cite{Giombi:2009,Sezgin:2003,Petkou:2003,Leonhardt:2002}.
\subsubsection{Fourier Transform}

However  \eqref{FT} shows that the Fourier transform of \eqref{erg30} has a factor $\Gamma(D-3)$ that cancels the same factor in the denominator, and we get a finite $k$ independent result in momentum space at D=3.

\[ 
\int d^Dx_1\int d^Dx_2\int d^Dx_3 e^{-ik_1.x_1 -ik_2x_2-ik_3x_3}\frac{1}{(x_1-x_2)^{2}(x_3-x_1)^{2}(x_1-x_3)^{2}}
\]
\[\boldsymbol =
\int d\alpha_1 \int d\alpha_2\int d\alpha_3\delta(\alpha_1+\alpha_2+\alpha_3-1)\alpha_1^{\Dt-2}\alpha_2^{\Dt-2}\alpha_3^{\Dt-2}
\]
\be 
\times \frac{\Gamma(D-3)}{(k_1^2 \alpha_2\alpha_3+k_2^2 \alpha_1\alpha_3+k_3^2 \alpha_2\alpha_1)^{D-3}}
\ee

Plugging this into \eqref{erg30} we see that in $D=3$ the momentum dependence disappears.
\subsubsection{Momentum Space}

A direct momentum space calculations Appendix \eqref{B} gives the same result.
The propagators  inside the loop are $\frac{1}{k^2}$ being evaluated at $\lm=0$. In this situation, in $D=3$ the loop integral becomes $\frac{1}{k_1k_2k_3}$ as given in Appendix \eqref{B}  and ${\cal G}(k)$ is $k$ as given in Appendix \eqref{A}. Thus we get a momentum independent constant:
\be
S_{cl}=\frac{2g}{3}~  \int_{k_1,k_2,k_3}\dd(k_1+k_2+k_3)J(k_1)J(k_2)J(k_3)
\ee
This is the result stated in the last subsection. The position space answer is zero and the momentum space answer is finite momentum independent. There is no inconsistency:
This is analytic in momenta and in fact corresponds, in position space, to a local term in the bulk action for $J$ and thus can be renormalized away. So in fact the correlator can be set to zero.  Cancellation of zeroes between numerator and denominator makes the answer an  indeterminate (constant).

This concludes the discussion of the ERG equation and its holographic formulation. We now turn to map the bulk action to AdS space following \cite{Sathiapalan:2017,Sathiapalan:2019}.

\section{Mapping to $AdS$ space}

In the last section we obtained a holographic formulation but the space was not AdS. As shown in \cite{Sathiapalan:2017} it is
possible to do a field redefiniton that maps this theory to a scalar field theory in $AdS_{D+1}$ . We take the Poinacre
patch metric:
\[
ds^2=\frac{dz^2+dx^idx_i}{dz^2}
\]
\subsection{Mapping}

We now map to AdS space using the techniques of \cite{Sathiapalan:2017} reviewed in Section \eqref{rev}. The first step is to do a field redefinition from $J(p,z)$ to $y(p,z)$
given by
\be
J=fy
\ee
where $f$ is a scale dependent rescaling and
it was shown there that it takes the form
\be
\frac{1}{f(p,z)}=z^{\Dt}(A(p)K_\nu(pz) + B(p)I_\nu(pz))
\ee
with $\nu=\frac{1-\dd}{2}$ where $D=3+\dd$. 
And $G(p,z)$ takes the form
\be
G(p,z)=\frac{C(p)K_\nu(pz)+D(p)I_\nu(pz)}{A(p)K_\nu(pz) + B(p)I_\nu(pz)}
\ee
$A,B,C,D$ are constants that are to be fixed by imposing suitable boundary conditions and satisfy $AD-BC=1$.

We note the following asymptotes for the Bessel functions:
\br
K_\nu(pz)&\to & \hf(\frac{2}{pz})^{\nu}\Gamma(\nu)~~~pz \to  0\nonumber \\
K_\nu(pz)&\to &\sqrt{\frac{\pi}{2}}\frac{e^{-pz}}{\sqrt {pz}}~~~pz\to \infty \nonumber\\
I_\nu(pz)&\to &(\frac{pz}{2})^\nu\frac{1}{\Gamma(1+\nu)}~~~~pz\to 0\nonumber\\
I_\nu(pz)&\to & \frac{e^{pz}}{\sqrt{2\pi pz}}~~~~pz\to  \infty
\er
We would like $G(\infty)=0$ and $G(0)\approx p^{-2\nu}$.
Clearly $D=0$ if $G$ has to vanish at infinity. This gives 
\be
BC=-1
\ee
As $pz\to 0$ we have
\be   \label{g}
G\to \frac{C(p)}{A(p)}= \gamma p^{-2\nu}
\ee
where $\gamma$ is the overall normalization of $G$.
We also require that $f$ should become a $p$-independent constant at the boundary $z=0$ so that it doesn't modify the behaviour of the Green function and also to ensure that $J(p)$ and $y(p)$ are essentially the same external sources in the boundary field theory. 
\[
f(p,z)\to z^{-\Dt}(\frac{pz}{2})^{\nu}\frac{1}{A(p)\Gamma(\nu)}~~as~~~pz\to 0
\]
So we choose 
\be  \label{a}
A(p)= p^\nu \implies f\to z^{-\Dt+\nu}
\ee
From \eqref{g} and \eqref{a} we get
\be
C(p)=\gamma p^{-\nu}~~~~B(p)= -\frac{1}{\gamma}p^\nu
\ee
Thus we get
\br  \label{f2}
\frac{1}{f(p,z)}&=& z^\Dt p^\nu (K_\nu(pz)-\frac{1}{\gamma}I_\nu(pz))\\
 G(p,z)&=&  \frac{\gamma p^{-\nu}K_\nu(pz)}{ p^\nu (K_\nu(pz)-\frac{1}{\gamma}I_\nu(pz))}\label{g2}\\
 \frac{f}{G}&=& z^{-\Dt}\frac{p^\nu}{\gamma K_\nu(pz)} \label{fg2}
 \er

Note that when $D=3$, $\nu=\hf$. In that case
\br
G(z,p)&=& \frac{\frac{\gamma}{\sqrt p}K_\hf (pz)}{\sqrt p(K_\hf(pz) -\frac{1}{\gamma}I_\hf(pz))} \\
\frac{1}{f(z,p)}&=& z^\Dt\sqrt p (K_\hf(pz) -\frac{1}{\gamma} I_\hf(pz))
\er

Also
\[
K_\hf(pz)= \frac{e^{-pz}}{\sqrt{pz}}\sqrt \frac{\pi}{2},~~~~I_\hf(pz)= \sqrt\frac{2}{\pi}\frac{sinh(pz)}{\sqrt{pz}}
\]
are exact expressions.

\subsection{Kinetic Term}

The mapping is chosen \cite{Sathiapalan:2017} so that the kinetic term has the standard AdS form
\be   \label{k}
\int_p \int dz z^{-D+1}\{(\p_z y(p) \p_z y(-p) + (p^2 + \frac{m^2}{z^2})y(p)y(-p)\}
\ee
with $\nu^2=m^2+\frac{D^2}{4}$. And the scaling dimensions of the boundary operators are $\DD^\pm = \Dt \pm \nu$. Using $\nu-\frac{1-\dd}{2}$ gives $\DD^+=2$ and $\DD^-=D-2$.  In terms of $\dd$ we have
\be
m^2=-2(1+\dd)
\ee 
So for $D=3$, $m^2=-2$.  $\nu=\hf$ so the dimensions are $\DD^+=2$ and $\DD^-=1$. In our case for $D=3$, $\chi$ has scaling dimension 2 in the boundary theory and the dual source $J$ thus has dimension 1.

Here $y(x,z)$ is dimensionless (i.e. by multiplying with appropriate powers of the cutoff $\lm$) and $y(p,z)$ therefore has dimension $-D$. 

Note that there is a also a boundary term contribution to $S^{Bulk}_{kinetic}$ at $z=\infty$ and at  $z=\eps \to 0$ \cite{Sathiapalan:2017}:
\be   \label{boundf}
\hf (z^{-D+1}y^2\frac{d \ln f}{dz})\Big|_{z=\eps}^{z=\infty}
\ee
Since the solutions we consider vanish at $z=\infty$ there is no contribution from there. At $z=0$, we have seen in \eqref{f2} that $f\approx z^{-\Dt+\nu}$. Thus the contribution at $z=0$ is a $p$-independent constant. We will neglect this because this does not contribute a non analytic $p$-dependence to the Green function and hence can be renormalized away.

\subsection{Cubic Term}

The cubic term in the bulk action is given in \eqref{evop1}:
\[
S_{int}=\frac{1}{\sqrt N}\int_{k_1,k_2,k_3}\dd(k_1+k_2+k_3)f(k_1,k_2,k_3,t)J(k_1)J(k_2)J(k_3)
\]
 \be   
=
\frac{1}{\sqrt N}\int_0^\infty dt~  \int_{k_1,k_2,k_3}\dd(k_1+k_2+k_3)\int _p \frac{d}{dt} (\DD_h(p)\DD_h(p+k_1)\DD_h(p+k_1+k_2)) \frac{1}{ G(k_1,t)G(k_2,t)G(k_3,t)} J_0(k_1,t)J_0(k_2,t)J_0(k_3,t) 
\ee
to leading order.

Writing $J=fy$ we get
\be
S_{int}=
\frac{1}{\sqrt N}\int_0^\infty dt~  \int_{k_1,k_2,k_3}\dd(k_1+k_2+k_3)\int _p \frac{d}{dt} (\DD_h(p)\DD_h(p+k_1)\DD_h(p+k_1+k_2)) \frac{f(k_1,t)f(k_2,t)f(k_3,t)}{ G(k_1,t)G(k_2,t)G(k_3,t)} y(k_1,t)y(k_2,t)y(k_3,t) 
\ee

Let us write ($\lo$ is the UV cutoff, $\lm$ is the moving IR cutoff and $\lm=\lo e^{-t}$)
\be
I(k_1,k_2,k_3,\lm,\lo)=\int _p  (\DD_h(p)\DD_h(p+k_1)\DD_h(p+k_1+k_2))
\ee
Using \eqref{fg2} we get

\[
\tcboxmath{S_{int}^{Bulk}=
\frac{1}{\sqrt N}\int_{-\infty}^\infty dt~  z^{-3\Dt}\int_{k_1,k_2,k_3}\dd(k_1+k_2+k_3) \frac{d}{dt} I(k_1,k_2,k_3,\lm,\lo) \frac{(k_1)^\nu}{\gamma K_\nu(k_1z)}\frac{(k_2)^\nu}{\gamma K_\nu(k_2z)}\frac{(k_3)^\nu}{\gamma K_\nu(k_3z)}}
\]
\be   \label{actiony}
\tcboxmath{\times ~~y(k_1,t)y(k_2,t)y(k_3,t) }
\ee

It is difficult to write $\DD_h$ explicitly given the complicated form for $G$, so the integral cannot be evaluated exactly.
(We remind the reader that $[\DD_h (x - y)]^2 = {\cal G}(x - y)$ and ${\cal G}(p) = \frac{\gamma}{p} - G(p)$, so given a form for $G(p)$ in
\eqref{g2} one has to extract $\DD_h (x - y)$.)
So we leave the answer for the cubic term in the bulk action in this form.

Nevertheless one can argue as follows: $\DD_h$ is a high energy propagator, so $I(k_1 , k_2 , k_3 , \lm, \lo )$ contains
contributions from high energy modes between $\lm$ and $\lo$ - we also assume that $\lo\to \infty$ for convenience. The
derivative $\frac{dI}{d\lm}$ is the contribution from modes between $\lm$ and $\lm+d\lm$. This can also be calculated as the
derivative of the contribution of low energy modes from $0 - \lm$, i.e. $\lm$ plays the role of a UV cutoff in this
calculation. If we assume that for $k_i << \lm$, the leading term is universal and that the precise nature of the
cutoff is not important, one can calculate this analytically and obtain a closed form expression. This is done
in Appendix (\eqref{B}) (\eqref{I1}) :
\[ 
\tcboxmath{\lm\frac{dI(k_i,\lm )}{d\lm}= \frac{8}{(4\pi)^\Dt \Gamma(D-3)} ~(\frac{z}{2})^{2(D-3)}}  \times
\]
\be  \label{I1}
\tcboxmath{(\frac{2k_1}{z})^{-\nu}K_{\nu}(k_1z)(\frac{2k_2}{z})^{-\nu}K_{\nu}(k_2z)(\frac{2k_3}{z})^{-\nu}K_{\nu}(k_3z)}
\ee
(with $z=\frac{1}{\lm}$)

We can plug this into \eqref{actiony}. (The approximation $k_ i << \lm$ is also justified because the solutions $y(k_ i , z)$
that are used in calculations of correlations vanish for large $z$ or small $\lm$.) The momentum dependence cancels in the
factor multiplying $y(k_ 1 , z)y(k_ 2 , z)y(k_ 3 , z)$ and we get, in position space:

\be   \label{yyy}
S_{int}^{Bulk}=\frac{1}{\sqrt N}\frac{2^{3-\Dt}}{(4\pi)^\Dt\Gamma(D-3)\gamma^3}\int_{-\infty}^\infty dz~z^{-D-1}\int d^Dx~y(x,z)^3
\ee

Note that it vanishes when $D = 3$ just as in AdS/CFT calculations \cite{Giombi:2009,Sezgin:2003, Petkou:2003, Leonhardt:2002}.
\subsection{Correlation Functions}

We substitute a leading term in the solution of the EOM for $y(p,z)$ which is of the form
\be 
y(p,z)= Y(p)z^\Dt K_\nu(pz)
\ee
We have kept only the $K_\nu$ term because $I_\nu$ blows up at $z=\infty$. We choose $Y(p)$ so that
at $z=\eps$ 
\be
y(p,\eps)=J_c(p) \implies Y(p)=\frac{J_c(p)}{K_\nu(p\eps)}
\ee
Thus
\be   \label{y}
y(p,z)= J_c(p) \frac{ z^\Dt K_\nu(pz)}{\eps ^\Dt K_\nu(p\eps)}
\ee

\subsubsection{Two Point Function:}

Plugging the solution \eqref{y} into the kinetic term \eqref{k} gives the boundary term
\be
\int_p z^{-D+1}y\frac{d y}{dz}\Big|_{z=0}^{z=\infty}
\ee
in addition to the boundary contribution \eqref{boundf} which we can neglect. 

This gives the form familiar from AdS/CFT calculations:
\be
W[J_c]=\int _p \eps^{-D}J_c(p)J_c(-p)\frac{\frac{zdK_\nu(pz)}{dz}\Big|_{z=\eps}}{K_\nu(p\eps)}
\ee

This can be evaluated for small $p\eps$ and one obtains the well known result \cite{Penedones2016}:
\be   \label{112}
W[J_c]=\int _p \eps^{-D}J_c(p)J_c(-p)\frac{\Gamma(1-\nu)}{\Gamma(1+\nu)}2\nu (\frac{p\eps}{2})^{2\nu}
\ee
where the non-analytic term is displayed. Any term that is analytic is a local term that can be changed by counterterms and thus has no physical significance. The propagator of $\chi$ thus behaves as $p^{2\nu}$ as expected on dimensional grounds.

\subsubsection{Three Point Function:}
Plugging this into \eqref{actiony} we get
Rather than plugging \eqref{y} into \eqref{yyy} it is easier to plug into \eqref{actiony} because we end up with a boundary
term:
\be   \label{erg31}
W_1[J_c]=-\frac{2}{3\gamma^3\sqrt N}\int_{-\infty}^\infty dt~\eps^{-3\Dt} \int_{k_1,k_2,k_3}\dd(k_1+k_2+k_3) \frac{d}{dt} I(k_1,k_2,k_3,\lm,\lo)\frac{(k_1k_2k_3)^{\nu}}{K_\nu(k_1\eps)K_\nu(k_2\eps)K_\nu(k_3\eps)}J_c(k_1)J_c(k_2)J_c(k_3)
\ee

A factor $\frac{2}{3}$ is due to a similar contribution from the kinetic term. All the $z$-dependence (other than in $I$)
cancels. This is not a coincidence, since both $y$ and $G/f$ obey the same equation with the same boundary
conditions. We are left with the boundary contributions at $z = 0$ and $z = \infty$. The contribution at $z = 0$
vanishes and we pick up the contribution at $z = \infty$ or $\lm = 0$. All this is of course the same as what we saw in
Section 3. So we get
\be 
W_1[J_c]=-\frac{2}{3\gamma^3\sqrt N}~\eps^{-3\Dt} \int_{k_1,k_2,k_3}\dd(k_1+k_2+k_3)  I(k_1,k_2,k_3,0,\lo)\frac{(k_1k_2k_3)^{\nu}}{K_\nu(k_1\eps)K_\nu(k_2\eps)K_\nu(k_3\eps)}J_c(k_1)J_c(k_2)J_c(k_3)
\ee

 So we are left with a UV cutoff $\lo$ which we can take as $1/\eps$. In limiting cases where we take $\lo\to \infty$  the details of the regulator are not important one can choose a simple regulator as is done in Appendix \eqref{B} \eqref{b138}. The integrals can bedone in $D=3$($\nu=\hf$).
 \be   
\tcboxmath{I(k_1,k_2,k_3,0,\lo)=
\frac{1}{(4\pi)^{\frac{D}{2}}} 2(k_1\lo)^{-\hf} K_{\hf}(\frac{2k_1}{\lo}) 2(k_2\lo)^{-\hf} K_{\hf}(\frac{2k_2}{\lo})
 2(k_3\lo)^{-\hf} K_{\hf}(\frac{2k_3}{\lo})}
\ee
In the limit $\lo\to \infty$ this becomes 
\be
I(k_1,k_2,k_3,0,\infty)= 2\sqrt 2\frac{1}{k_1k_2k_3}
\ee
and also when $\eps \to 0$
\[
\frac{(k_1k_2k_3)^{\nu}}{K_\nu(k_1\eps)K_\nu(k_2\eps)K_\nu(k_3\eps)}=
(k_1k_2k_3)^{2\nu}\frac{2\sqrt 2}{[\Gamma(\nu)]^3}\eps^{3\nu}
\]

and plugging all this into \eqref{erg30} and setting $\nu=\hf$ we get
\be
W_1[J_c]=- \frac{8}{\gamma^3\sqrt N}\int_{k_1,k_2,k_3}\dd(k_1+k_2+k_3)J_c(k_1)J_c(k_2)J_c(k_3)
\ee
Note that the two point function in \eqref{112} had a factor of $\eps^{-2}$ (in$ D = 3$ i.e. $\nu = \hf$ ) and the three point
function has $\eps^{-3}$ . Thus we can absorb these powers into a renormalization of $J_ c (k_ i )$ by setting $J_{ CR} = \frac{J_c}{\eps}$ and
the limit $\eps \to 0$ can be taken.

Thus we see that in $D = 3$  ($\dd = 0$) the dependence on external momenta vanishes. Thus the correlation is
a local function in three dimensions and {\em can be renormalized away}. As explained in Section 3 another way to
see this is that if we Fourier transform to k-space in D-dimensions, a conformal correlator, we get a factor of
$\Gamma(D-3)$ (see Appendix (\eqref{C})) and in 3 dimensions this factor diverges. Thus in three dimensions the coefficient of
the conformal correlator vanishes in x-space. We have seen this in the boundary calculation also. As mentioned
in the introduction, agreement with the boundary is guaranteed because we are just doing an ERG calculation
in a somewhat unusual (holographic) language - the language being motivated by the AdS/CFT correspondence.

\section{Summary and Conclusions}

In this paper an attempt has been made to extend the ideas of \cite{Sathiapalan:2017,Sathiapalan:2019} to the $O(N )$ vector model in $D$ dimensions
at the Wilson-Fisher fixed point. Instead of studying correlations of a fundamental scalar field we focus on
correlations of composite operators of the form $\phi^I\phi_I$. This is done by introducing an auxiliary field $\chi$ and an
action is obtained for $\chi$. The interaction terms are down by powers of 
$\sqrt N$ . An ERG equation for the generating
functional $Z[J]$ where $J$ is a source for$\chi$, is written down. The
 resulting equation is similar to Polchinski ERG
equation but contains higer order corrections as powers of $1/\sqrt N$  . This is a generalization of the usual Polchinski
ERG equation and plays an important role in this attempt to make contact with Holographic RG\cite{Giombi:2009,Sezgin:2003, Petkou:2003, Leonhardt:2002}.

The leading correction term can be approximated by a cubic term in $J$ and at higher orders there are
additional corrections. An evolution operator for $Z[J]$ can be written down as a functional integral of a $D + 1$
dimenional action for $J$. It has a kinetic term analogous to what was obtained in \cite{Sathiapalan:2017} but also additional potential
terms. This action can be mapped to an action in AdS space for a scalar field. The leading 
term is a kinetic
term and in addition there are potential terms that can be written down in a power series in $1/\sqrt N$ . The leading
cubic term has been calculated and vanishes in $ D = 3$ in agreement with earlier results\cite{Giombi:2009,Sezgin:2003, Petkou:2003, Leonhardt:2002}. Since
the bulk theory obtained here is mathematically equivalent to the ERG evolution operator by construction,
agreement between bulk and boundary is built into the formalism. The interesting question then is whether
the bulk theory obtained here is indeed the higher spin theory \cite{Vasiliev:2003,Vasiliev:2004} as suggested in \cite{Klebanov:2002,Sezgin:2003}. Calculation of
four and higher point functions in this approach would shed light on this question.

An obvious composite operator that can be studied is the energy momentum tensor. The source for this
is a spin two massless field which is the graviton of the bulk theory. One should thus obtain dynamical AdS
bulk gravity from any critical theory that has an energy momentum tensor. Furthermore since there an infinite
number of composite operators, the dual also has an infinite number of fields in addition to the metric tensor. It
would be interesting to study these operators in this approach. Finally these techniques can be applied to other
CFT’s in any dimension since ERG equations are very generally applicable. We leave these open questions for
the future.

\vspace{1cm}

{\bf Acknowledgements:} I would like to thank H. Sonoda for many useful discussions.

\begin{appendices}

\section{Kinetic Term for $  \chi$} \label{A}
\subsection{Evaluating the one loop diagram with high energy propagator}

The integral we need is
\[
I(k)=\int \Dp \DD_h(p)\DD_h(k+p)
\]
\[
=\int \Dp \frac{1-e^{\frac{(k+p)^2}{\lm^2}}}{(k+p)^2}\frac{1-e^{\frac{p^2}{\lm^2}}}{p^2}
\]
\[
= \int_0^\infty ds_1 \int _0^\infty ds_2 \int \Dp e^{-s_1 p^2 -s_2 (k+p)^2}(1-e^{\frac{(k+p)^2}{\lm^2}})(1-e^{\frac{p^2}{\lm^2})}
\]
Let $\frac{1}{\lm^2}=a$. Then we have integrals of the form
\[
I(k,x,y)= \int_0^\infty ds_1 \int _0^\infty ds_2 \int \Dp e^{-(s_1+x) p^2 -(s_2+y) (k+p)^2}
 \]
 were $x,y$ can be either $0$ or $a$. 

Doing the $p$ integral gives
\[
I(k,x,y)= \frac{1}{(4\pi)^{\frac{D}{2}}} \int_0^\infty ds_1 \int _0^\infty ds_2 \frac{1}{(s_1+x+s_2+y)^\Dt}e^{-k^2\frac{(s_1+x)(s_2+y)}{s_1+x+s_2+y}}
\]
Let $z_1=s_1+x$ and $z_2=s_2+y$.
\[
I(k,x,y)=\frac{1}{(4\pi)^{\frac{D}{2}}} \int_x^\infty ds_1 \int _y^\infty ds_2 \frac{1}{(z_1+z_2)^\Dt}e^{-k^2\frac{z_1z_2}{z_1+z_2}}
\]
Perform the usual change of variables:
\[
z_1=\alpha_1t,~z_2=\alpha_2 t,~~~ \alpha_1+\alpha_2=1,~~~z_1+z_2=t
\] 
The measures are related by
\[
dz_1dz_2= d\alpha_1 dt~t
\]
\be   \label{a0}
I(k,x,y)=\frac{1}{(4\pi)^{\frac{D}{2}}}\int_0^1 d\alpha_1 \int_{x+y}^\infty  dt~ t^{1-\Dt} e^{-k^2 \alpha_1\alpha_2 t}
\ee
\be   \label{a1}
I(k)=I(k,0,0)+I(k,a,a)-I(k,a,0)-I(k,0,a)
\ee
We evaluatethe integral $I(k,x,y)$ in some limits.

\begin{enumerate}
\item {${\boldsymbol \lm =\infty}$}
The high energy propagator vanishes. In this case $I(k)=0$ since $a=0$ and the terms in \eqref{a1} cancel.

\item{$\mathbf \lm=0$} The high energy propagator reduces to the ordinary one $\frac{1}{p^2}$. $a=\infty$ and so  $I(k)=I(k,0,0)$. Now there is no IR cutoff so we expect  the integralto diverge as $k\to 0$.
\be
I(k,x,y)=\frac{1}{(4\pi)^{\frac{D}{2}}}\int_0^1 d\alpha_1 \int_0^\infty  dt~ t^{1-\Dt} e^{-k^2 \alpha_1\alpha_2 t}
\ee

 Change  variables to $\beta,t$ where $\alpha_1 \alpha_2 t= \beta$. Note that $t>4\beta$.  We get a measure

\[
\int d\alpha_1 dt =\int~ d\beta ~dt \frac{1}{t(\sqrt{1-\frac{4\beta}{t}})}
\]
Let $\frac{4\beta}{t}=\bar t$. Then
\[
I(k,0,0)= \frac{1}{(4\pi)^{\frac{D}{2}}}\int _0^\infty d\beta \int_0^1 d\bar t (\bar t)^{\Dt} (1-\bar t)^{-\hf} (4\beta)^{-\Dt +1} e^{-k^2\beta -\frac{1}{\beta \lo^2}}
\]
A  cutoff $\lo$ has been introduced to regulate the UV end ($\beta\to \infty$)of the  integral. Actually for the case at hand $D < 4$ the
integrals are UV convergent and we can take $\lm \to \infty$ if we want. 
Using
\[
\int _0^\infty d\beta \beta ^{-\nu -1}e^{-k^2\beta -\frac{1}{\beta \lm^2}}= 2(k\lm)^\nu K_\nu(\frac{2k}{\lo})
\] we get setting $D = 4 - 2\eps$, for the case where $\lm = 0$:
\be
I_{\lo}(k,0,0)=  \frac{1}{(4\pi)^{\frac{D}{2}}}\frac{\Gamma (\Dt+1)\Gamma(\hf)}{\Gamma(\frac{D+3}{2})} 2(k\lo)^{-\eps} K_\eps(\frac{2k}{\lo})
\ee
The limit $\lo\to \infty$ can be taken as long as $\eps > 0$ and we get
\[
I(k, 0, 0) \approx k^{-2\eps}
\]
This is non analytic in $k$ at $k = 0$. In \eqref{a0} it can be seen that when $k^2  = 0$ the $t \to \infty$ region of
integration is divergent. So although the integrand looks analytic in $k^2$ , the integral is not. On the other
hand in $I(k)$ (\eqref{a1}) there are cancellations of the divergences at the $t = \infty$ region and we can expect analyticity.

\item{\bf$\mathbf \lm$ finite}
We will expand $I(k)$ in powers of$ k^ 2$ and the result is a sum of finite terms as we see below. Go back to \eqref{a0}. Expand the exponential:
\be
I(k,x,y)=\frac{1}{(4\pi)^{\frac{D}{2}}}\int_0^1d\alpha_1\int_{x+y}^\infty dt~
t^{1-\Dt}(1-k^2 \alpha_1\alpha_2 t+ \frac{1}{2!}(k^2 \alpha_1\alpha_2 t)^2-...+(-1)^n\frac{1}{n!}(k^2 \alpha_1\alpha_2 t)^n+...)
\ee

Write
\be
I_n(0,x,y)=\frac{1}{(4\pi)^{\frac{D}{2}}}\int_0^1d\alpha_1\int_{x+y}^\infty dt~
t^{1-\Dt}(-1)^n( \alpha_1\alpha_2 t)^n
\ee
This can be easily evaluated:
\[
=\frac{1}{(4\pi)^{\frac{D}{2}}}(-1)^n\frac{\Gamma(n+1)^2}{\Gamma(2n+2)}[\frac{T^{n+\eps}-(x+y)^{n+\eps}}{n+\eps}]
\]
The upper end of the $t$-integration has been cutoff at $T$ . This is the IR divergence due to vanishing $k^ 2$ .
Denoting the coefficient of $\frac{(k^2)^n} {n!}$ ) in $I(k)$ by $I_ n (0)$, we have
\[
I_n(0)=I_n(0,0,0)+I_n(0,a,a)-I_n(0,a,0)-I_n(0,0,a)
\]
The divergent terms in $T$ cancel and we are left with
\be
I_n(0)=(-1)^n\frac{1}{(4\pi)^{\frac{D}{2}}}\frac{\Gamma(n+1)^2}{\Gamma(2n+2)}\frac{1}{(\lm^2)^{n+\eps}}\frac{1}{n+\eps}(2-2^{n+\eps})
\ee
and we get a manifestly analytic power series in $k^2$:
\be
\tcboxmath{I(k)=\sum_n(-1)^n\frac{1}{(4\pi)^{\frac{D}{2}}}\frac{\Gamma(n+1)^2}{\Gamma(2n+2)}\frac{1}{n!}(\frac{k^2}{\lm^2})^n\frac{1}{(\lm^2)^{\eps}}\frac{1}{n+\eps}(2-2^{n+\eps})}
\ee
\subsection{Kinetic term for $\chi$ when $\lm = 0$}

 Thus, including the mass term, the quadratic part of the effective action for $  \chi$ is
\be   \label{s2chi}
S_2[  \chi]= \frac{1}{(4\pi)^{\frac{D}{2}}}\frac{\Gamma (\Dt+1)\Gamma(\hf)}{\Gamma(\frac{D+3}{2})}\int _k~  \chi(k) [2(k\lo)^{-\eps} K_\eps(\frac{2k}{\lo}) ]  \chi(-k)+\frac{1}{2\bar u} \int _k~ \chi(k)  \chi(-k) 
\ee

{\bf D=3:}
In this case $\eps=\hf$ and we have
\be
S_2[  \chi]= \frac{1}{(4\pi)^{\frac{3}{2}}}\frac{\Gamma (\frac{5}{2})\Gamma(\hf)}{\Gamma(3)}\int _k~  \chi(k) [2(k\lo)^{-\hf} K_\hf(\frac{2k}{\lo}) ]  \chi(-k)+\frac{1}{2\bar u} \int _k~ \chi(k)  \chi(-k) 
\ee
\end{enumerate}

We use the asymptotic series for $\frac{k}{\lo}\to 0$:
\[
(2k\lo)^{-\eps}K_\eps(\frac{2k}{\lo})= \frac{2^\eps}{\lo^{2\eps}}[2^{-1-\eps} \Gamma(-\eps)+...]+ 2^{-\eps}k^{-2\eps}[2^{-1+\eps} \Gamma(\eps) + O((\frac{2k}{\lo})^2)]
\]
%

Choosing
\be
 \bar u= \frac{\lo^{2\eps}}{\Gamma(-\eps)}\frac{\Gamma(\frac{D+3}{2})}{\Gamma (\Dt+1)\Gamma(\hf)}(4\pi)^{\frac{D}{2}} \equiv \bar u^*
\ee
  we get a conformal theory.
Note that $\bar u^*\to \infty$ as $\lo \to \infty$. This is the expected fixed point value for finite values of $\eps$. For the critical theory
\[
S_2[  \chi]= \frac{1}{(4\pi)^{\frac{D}{2}}}\frac{\Gamma (\Dt+1)\Gamma(\hf)}{\Gamma(\frac{D+3}{2})} 2^{-1-\eps}\Gamma(\eps)\int _k   \chi(k)[   k^{-2\eps} ]  \chi(-k)
\]
{\bf D=3:} In $D=3,~ \eps=\hf$ we know that the fixed point large $N$ action has $\bar u\to \infty$. We see that the action for $S_2$:
\be
S_2[  \chi] \approx   \int _k   \chi(k) k^{-1}   \chi(-k)
\ee
has the conformally invariant form. Dimension of $  \chi(k)$ is -1. Dimension of $  \chi(x)$ is $2$. (This last statement is true for any $\eps$.) We also see from
\eqref{z0} that $  \chi^2$ being irrelevant in $D<4$, $  \chi$ imposes a constraint $\phi^I\phi^I=\frac{r}{2\bar u}$, thus making it a non linear sigma model.

\begin{figure}
\hspace{4cm}\includegraphics[width=8cm]{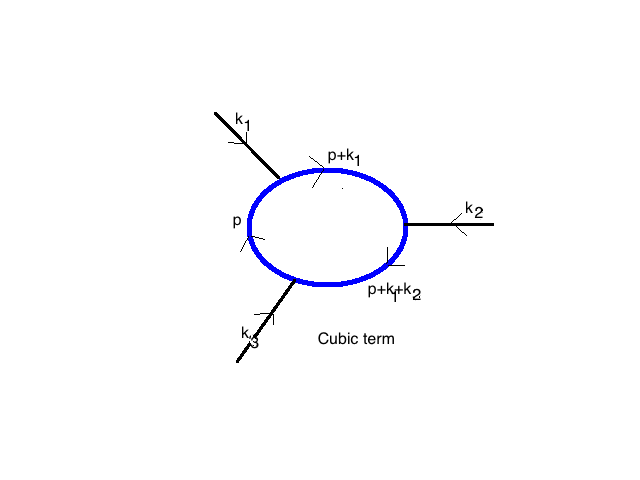}
\caption{Cubic term generating $\chi^3$ interaction. The internal lines are $\phi$ propagators $\DD_h$.}
\end{figure}

\section{Cubic Term}\label{B}
The quantity that needs to be evaluated is $\DD_{hx'y'}\DD_{hx'z'}\DD_{hz'y'}$ corresponding to the one loop diagram in Figure
5. We evaluate it in momentum space. We cannot evaluate this for the cutoff function that is demanded by the
map to AdS space because it is very complicated. However when $\frac{k}{\lm}\to 0$, in a critical theory, one expects the
leading terms to be universal and in fact determined by the scaling dimensions. So we will use some convenient
forms of the cutoff in this section.

\subsection{$\lm\to 0$ limit}
\begin{enumerate}

\item The Feynman diagram is ($k_1+k_2+k_3=0$).
\be  \label{start}
I(k_1,k_2,k_3,0,\lo)=\int \Dp \frac{1}{(k_1+p)^{2a_3}}\frac{1}{(k_1+k_2+p)^{2a_1}}\frac{1}{(p)^{2a_2}}
\ee
with some regulator $\lo$. The regularization scheme  we leave unspecified for now. We are eventually interested in $a_i=1$ and $D=3$.

So  effective action at the cubic order is (assume $a_i=a$)
\[
S_3=\frac{1}{\sqrt N}\int_{k_1}\int_{k_2}\int_{k_3}\chi(k_1)\chi(k_2)\chi(k_3)I(k_1,k_2,k_3,\lm)(2\pi)^D\delta^D(k_1+k_2+k_3)
\]

\item Thus we need
\be
I=\int ds_1\int ds_2\int ds_3 \frac{s_1^{a_1-1}s_2^{a_2-1}s_3^{a_3-1}}{\Gamma(a_1)\Gamma(a_2)\Gamma(a_3)}\int \Dp e^{-(k_1+p)^2s_3}e^{-(k_1+k_2+p)^2s_1}e^{-(p)^2s_2}
\ee

\item Simplify exponent. Use $k_1+k_2=-k_3$ and do the $p$ integrals to get

%
%
%
%
\be \label{17k}
I=\frac{1}{(4\pi)^{\frac{D}{2}}\Gamma(a_1)\Gamma(a_2)\Gamma(a_3)}\int ds_1\int ds_2\int ds_3 \frac{s_1^{a_1-1}s_2^{a_2-1}s_3^{a_3-1}}{(s_1+s_2+s_3)^{\frac{D}{2}}}
e^{-\frac{k_1^2s_2s_3+k_3^2s_1s_2+k_2^2s_1s_3}{(s_1+s_2+s_3)}}
\ee

\item Change of variables:
\[
s_1=\alpha_1 t,~~s_2=\alpha_2 t,~~s_3=\alpha_3 t,~~~~s_1+s_2+s_3=t,~~~\alpha_1+\alpha_2+\alpha_3=1
\]
\[ 
\int ds_1ds_2ds_3~=\int dt ~t^2 d\alpha_1d\alpha_2 d\alpha_3 \delta(1-\alpha_1-\alpha_2-\alpha_3)
\]

\item Further change of variables from $\alpha_1,\alpha_2,t\to \beta_1 \beta_2, \beta_3$

\be   \label{cov2}
\alpha_1 \alpha_2t=\beta_3,~~\alpha_1 \alpha_3t=\beta_2,~~\alpha_3 \alpha_2t=\beta_1
\ee
with Jacobian
\[
J=(\beta_2\beta_3+\beta_1\beta_3+\beta_2\beta_1)=\alpha_1\alpha_2\alpha_3 t^2
\]
\be \label{I}
I=\frac{1}{(4\pi)^{\frac{D}{2}}\Gamma(a_1)\Gamma(a_2)\Gamma(a_3)}\int d\beta_1d\beta_2d\beta_3 J^{a_1+a_2+a_3-D} \beta_1^{\frac{D}{2}-a_1-1}\beta_2^{\frac{D}{2}-a_2-1}\beta_3^{\frac{D}{2}-a_3-1}e^{-k_1^2 \beta_1 -k_2^2 \beta_2 -k_3^2 \beta_3}
\ee


\item Regularization: We need to regularize \eqref{I}. Clearly $\beta_i\to 0$ is the UV region. So let us introduce factors of $e^{-\frac{1}{\lo^2\beta_i}}$ in the integral.

So
\[
\tcboxmath{
I(k_1,k_2,k_3,0,\lo)=
\frac{1}{(4\pi)^{\frac{D}{2}}\Gamma(a_1)\Gamma(a_2)\Gamma(a_3)}\int d\beta_1d\beta_2d\beta_3 J^{a_1+a_2+a_3-D} \beta_1^{\frac{D}{2}-a_1-1}\beta_2^{\frac{D}{2}-a_2-1}\beta_3^{\frac{D}{2}-a_3-1}}\times
\]
\be \label{Ifinal}
\tcboxmath{
e^{-k_1^2 \beta_1 -k_2^2 \beta_2 -k_3^2 \beta_3-\frac{1}{\lo^2}(\frac{1}{\beta_1}+\frac{1}{\beta_2}+\frac{1}{\beta_3})}}
\ee
\item 
{\bf Assume $\mathbf {a_1+a_2+a_3-D=0}$}

Do $\beta$ integrals.
\be
\int_0^\infty d\beta ~\beta^{-\nu-1} e^{-k^2 \beta - \frac{1}{\beta \lo^2}}= 2(k\lo)^\nu K_\nu(\frac{2k}{\lo})
\ee
.


Then using  $\dd= 1-2\nu$ and $a_i=1, D=3+\dd$ this can also be written as
\be
I(k_1,k_2,k_3,0,\lo)=
\frac{1}{(4\pi)^\frac{D}{2}\Gamma(a_1)\Gamma(a_2)\Gamma(a_3)} 2(k_1\lo)^{\nu-1} K_{\nu-1}(\frac{2k_1}{\lo}) 2(k_2\lo)^{\nu-1} K_{\nu-1}(\frac{2k_2}{\lo})
 2(k_3\lo)^{\nu-1} K_{\nu-1}(\frac{2k_3}{\lo})
\ee

Let us set $a_i=1$ and $D=3$. Then $\nu = \hf$ and we get ($K_\nu=K_{-\nu}$)
\be   \label{b138}
\tcboxmath{I(k_1,k_2,k_3,0,\lo)=
\frac{1}{(4\pi)^{\frac{D}{2}}} 2(k_1\lo)^{-\hf} K_{\hf}(\frac{2k_1}{\lo}) 2(k_2\lo)^{-\hf} K_{\hf}(\frac{2k_2}{\lo})
 2(k_3\lo)^{-\hf} K_{\hf}(\frac{2k_3}{\lo})}
\ee
For small momenta or when $\lo\to \infty$ this becomes 

\be
I(k_1,k_2,k_3,0,\infty)=\frac{2\sqrt 2}{k_1k_2k_3}
\ee
This corresponds to a term in the effective action for $\chi$:
\[
S_3[\chi]=\frac{1}{3!}\int _{k_1}\int_{k_2}\int_{k_3}\frac{2\sqrt 2}{k_1k_2k_3}\chi(k_1)\chi(k_2)\chi(k_3)\dd(k_1+k_2+k_3)
\]
Given that $[\chi(k)]=-1$, this is a scale invariant expression.

\item
Let us rewrite \eqref{I} in a form that makes the $\lm$-derivative easy to calculate by using
\[
\int _0^\infty dx~\frac{x^{a-1}}{\Gamma(a)}e^{-x \frac{J}{\beta_1\beta_2\beta_3}}= (\frac{J}{\beta_1\beta_2\beta_3})^{-a}=(\frac{1}{\beta_1}+\frac{1}{\beta_2}+\frac{1}{\beta_3})^{-a}
\]

Write $4x=z^2$ and do each of the $\beta$ integrals to get

\[
\tcboxmath{I(k_i,\lm)=\frac{16}{(4\pi)^{\frac{D}{2}}\Gamma(D-a_1-a_2-a_3)\Gamma(a_1)\Gamma(a_2)\Gamma(a_3)}\int _{\frac{1}{\lm}}^\infty dz~z^{2(D-a_1-a_2-a_3)-1}}\times
\]
\be
\tcboxmath{(\frac{k_1}{z})^{\Dt-a_2-a_3}K_{\Dt-a_2-a_3}(2zk_1)(\frac{k_2}{z})^{\Dt-a_1-a_3}K_{\Dt-a_1-a_3}(2zk_2)(\frac{k_3}{z})^{\Dt-a_2-a_1}K_{\Dt-a_2-a_1}(2zk_3)}
\ee
where a UV regularization has been done.

\item Let us evaluate this in the following situation: $D=3+\dd$ and $a_i=1$.
So
\[
D-a_1-a_2-a_3=\dd ~~;~~~\Dt -2= -\hf +\frac{\dd}{2}~~~;\nu = \Dt-a_2-a_3= -\hf +\frac{\dd}{2}< 0
\]
\[ 
\tcboxmath{I(k_i,\lo )= \frac{4}{(4\pi)^\Dt \Gamma(D-3)}\int _{\frac{1}{\lm}}^\infty dz~(\frac{z}{2})^{2(D-3)-1}}  \times
\]
\be
\tcboxmath{(\frac{2k_1}{z})^{-\nu}K_{\nu}(k_1z)(\frac{2k_2}{z})^{-\nu}K_{\nu}(k_2z)(\frac{2k_3}{z})^{-\nu}K_{\nu}(k_3z)}
\ee

\[
K_{-\nu}(z)=K_\nu(z) = \frac{z^\nu}{2^{1+\nu}} [ \Gamma[-\nu] + O(z^2)] +  \frac{z^{-\nu}}{2^{1-\nu}} [ \Gamma[\nu] + O(z^2)] ~~~; z\to 0
\]
\[
K_{-\nu}(z)=K_\nu(z)= e^{-z} (\sqrt{\frac{\pi}{2 z}}+ O(\frac{1}{z\sqrt z}))~~~; z\to \infty
\]

So
\[
(\frac{k}{z})^\nu K_\nu(2zk) = (\frac{k}{z})^{-\hf +\frac{\dd}{2}}[ \frac{(2zk)^{-\hf +\frac{\dd}{2}}}{2^{\hf +\frac{\dd}{2}}}\Gamma[\hf -\frac{\dd}{2}]+...]=\hf k^{-1+\dd} \Gamma[\hf -\frac{\dd}{2}]
\]

We will evaluate $\lm \frac{dI}{d\lm}$.
  
\[ 
\tcboxmath{\lm\frac{dI(k_i,\lm )}{d\lm}= \frac{8}{(4\pi)^\Dt \Gamma(D-3)} ~(\frac{z}{2})^{2(D-3)}}  \times
\]
\be  \label{I1}
\tcboxmath{(\frac{2k_1}{z})^{-\nu}K_{\nu}(k_1z)(\frac{2k_2}{z})^{-\nu}K_{\nu}(k_2z)(\frac{2k_3}{z})^{-\nu}K_{\nu}(k_3z)}
\ee
(with $z=1/\lm$).

\be
\tcboxmath{\lm\frac{dI(k_i,\lm )}{d\lm}\approx \frac{2^{-2\dd}}{(4\pi)^\Dt}\frac{[\Gamma(2-\Dt)]^3}{\Gamma(D-3)}\lm^{-2\dd}(k_1k_2k_3)^{\dd-1}}
\ee

Note that $I$ as well as$\frac{dI}{dt}$  vanishe as $\dd\to 0$, i.e. as $D\to 3$.

%

\subsection{$\lm$ finite - analytic power series in $k_i^2$}
The integral that needs to be done is, as before

\be  
I(k_1,k_2,k_3,\lm)=\int \Dp \frac{1-K(k_1+p)}{(k_1+p)^{2a_3}}\frac{1-K(p-k_3)}{(-k_3 p)^{2a_1}}\frac{1-K(p)}{(p)^{2a_2}}
\ee
where we take $K(p) = e^{ - \frac{p^2}{\lm^2}}$. It is fairly obvious that this integral is analytic in $k_i$ because only modes between $\lm$ and $\lo$ are being integrated. Nevertheless for completeness we give the power series expansion.

Define
\be   \label{b146}
I(k_i,x,y,z)=\int ds_1\int ds_2\int ds_3 \int \Dp e^{-(k_1+p)^2s_3}e^{-(k_1+k_2+p)^2s_1}e^{-(p)^2s_2} e^{-(k_1+p)^2z}e^{-(k_1+k_2+p)^2x}e^{-(p)^2y}
\ee

with $x, y, z$ being either $0$ or $a =\frac{1}{\lm^2}$. 
Thus

\[
I(k_i ) = I(k_ii , 0, 0, 0) + I(k_i , 0, a, a) + I(k_i , a, a, 0) + I(k_i , a, 0, a)
\]
\be \label{ikxyz}
-(I(k_i , a, a, a) + I(k_i , 0, 0, a) + I(k_i , a, 0, 0) + I(k_i , 0, a, 0))
\ee
Doing the $p$ integral in \eqref{b146} gives
\[
I=\frac{1}{(4\pi)^{\frac{D}{2}}}\int dz_1\int dz_2\int dz_3 
e^{-\frac{k_1^2z_2z_3+k_3^2z_1z_2+k_2^2z_1z_3}{(z_1+z_2+z_3)}}
\]
where $z_1=s_1+x$,
 $z_2=s_2+y$ $z_3=s_3+z$.
 
Now change variables:
\[
z_1=\alpha_1t,~~~ z_2=\alpha_2t,~~~z_3=\alpha_3t,~~~z_1+z_2+z_3=t, \alpha_1+
 \alpha_1+ \alpha_1+=1
 \] 
The measures are related by:
\[
dz_1dz_2dz_3=dt~t^2 d\alpha_1d\alpha_2
\] 
Now expand the exponentials to get a power series in $k_i^ 2$ : 
\[
I(k_i,x,y,z)=\sum_{m,n,p}(-1)^{m+n+p}\frac{(k_1^2)^m}{m!}\frac{(k_2^2)^n}{n!}\frac{(k_3^2)^p}{p!}
\]
\[
\int_0^1d\alpha_1\int_0^1d\alpha_2\int_0^1d\alpha_3~\dd(\alpha_1+\alpha_2+\alpha_3-1)(\alpha_2\alpha_3)^m(\alpha_1\alpha_3)^n(\alpha_2\alpha_1)^p
\int_{x+y+z}^\infty dt~t^{2-\Dt+m+n+p}
\]
The $\alpha_i$ integrals give:
\be  \label{cmnp}
B(m + p + 1, m + 2n + p + 2)B(m + n + 1, p + n + 1) \equiv C(m, n, p)
\ee
The $t$ -integral gives
\[
\int_{x+y+z}^\infty dt~t^{2-\Dt+m+n+p}=\frac{T^{3-\Dt+m+n+p}-(x+y+z)^{3-\Dt+m+n+p}}{3-\Dt+m+n+p}
\]
$T$ is a cutoff for the $t \to \infty$ end. As before from \eqref{ikxyz} we see that the $T$ -dependence drops out. 
\[
\tcboxmath{I(k_i,x,y,z)=\sum_{m,n,p}\frac{(-1)^{m+n+p}}{m!n!p!}(\frac{k_1^2}{\lm^2})^m(\frac{k_2^2}{\lm^2})^n(\frac{k_3^2}{\lm^2})^p}\times
\]
\be
\tcboxmath{3C(m,n,p) [-(2)^{\eps+1+m+n+p} +(3)^{\eps+m+n+p}+1]\frac{1}{\eps+1+m+n+p}}
\ee
 with $C(m,n,p)$ defined in \eqref{cmnp} and $D=4-2\eps$ as before.

\end{enumerate}

\section{Fourier Transform to Momentum Space }\label{C}
The main point to note is the factor of $\Gamma(D - 3)$ that emerges in the Fourier transform.

FT of 
\[
\frac{1}{(x_1-x_2)^{2d_3}(x_3-x_1)^{2d_2}(x_2-x_3)^{2d_1}}
\]
is
\[
\int d^Dx_1\int d^Dx_2\int d^Dx_3 e^{-ik_1.x_1 -ik_2x_2-ik_3x_3}\int ds_1s_1^{d_1-1}\int ds_2s_2^{d_2-1}\int ds_3s_3^{d_3-1}
\]
\[e^{-s_3(x_2-x_1)^2-s_2(x_3-x_1)^2-s_1(x_2-x_3)^2}
\]

One can set $x_3=0$ using translational invariance and do the integrals over $x_1,x_2$ to get

a determinant factor
\[
[\frac{\pi}{4(s_1s_2+s_1s_3+s_2s_3)}]^{\Dt}
\]
And the exponent is
\be
-\frac{k_1^2s_1+k_2^2s_2+k_3^2s_3}{4(s_1s_2+s_2s_3+s_3s_1)}
\ee

So we have
\be
I= \int ds_1 \int ds_2 \int ds_3~[\frac{\pi}{4(s_1s_2+s_1s_3+s_2s_3)}]^{\Dt}s_1^{d_1-1}s_2^{d_2-1}s_3^{d_3-1}e^{-\frac{k_1^2s_1+k_2^2s_2+k_3^2s_3}{4(s_1s_2+s_2s_3+s_3s_1)}}
\ee

Now let $s_i=\frac{1}{t_i}$ and $t_1+t_2+t_3=t$.
\[
\det = 4(s_1s_2+s_1s_3+s_2s_3)=\frac{4(t_1+t_2+t_3)}{t_1t_2t_3}
\]
\[I=
\int dt_1\int dt_2\int dt_3 [\frac{\pi}{4t}]^\Dt t_1^{\Dt -d_1-1}t_2^{\Dt -d_2-1}t_3^{\Dt -d_3-1}e^{-\frac{(k_1^2 t_2t_3+k_2^2 t_1t_3+k_3^2 t_2t_1)}{4t}}
\]
Now we repeat the two steps:
\[
t_i=\alpha_it;~~~~\alpha_1+\alpha_2+\alpha_3=1
\]

So 
\[
\int dt_1\int dt_2\int dt_3= \int_0^1 d\alpha_1 \int_0^1 d\alpha_2\int_0^1 d\alpha_3\delta(\alpha_1+\alpha_2+\alpha_3-1)\int dt ~t^2
\]
Substituting, we get
\[
\int d\alpha_1 \int d\alpha_2\int d\alpha_3\delta(\alpha_1+\alpha_2+\alpha_3-1)\int dt ~t^2
\]
\[
 [\frac{\pi}{4t}]^\Dt t^{3\Dt -d_1-d_2-d_3-3}\alpha_1^{\Dt-d_1-1}\alpha_2^{\Dt-d_2-1}\alpha_3^{\Dt-d_3-1}e^{-t(k_1^2 \alpha_2\alpha_3+k_2^2 \alpha_1\alpha_3+k_3^2 \alpha_2\alpha_1)}
\]
\[
=\int d\alpha_1 \int d\alpha_2\int d\alpha_3\delta(\alpha_1+\alpha_2+\alpha_3-1)\int dt~ t^{D-1-d_1-d_2-d_3}\alpha_1^{\Dt-d_1-1}\alpha_2^{\Dt-d_2-1}\alpha_3^{\Dt-d_3-1}e^{-t(k_1^2 \alpha_2\alpha_3+k_2^2 \alpha_1\alpha_3+k_3^2 \alpha_2\alpha_1)}\]

Do the $t$ integral to get:
\[ \tcboxmath{
\int d^Dx_1\int d^Dx_2\int d^Dx_3 e^{-ik_1.x_1 -ik_2x_2-ik_3x_3}\frac{1}{(x_1-x_2)^{2d_3}(x_3-x_1)^{2d_2}(x_1-x_3)^{2d_1}}}\]
\[\tcboxmath{\boldsymbol =
\int d\alpha_1 \int d\alpha_2\int d\alpha_3\delta(\alpha_1+\alpha_2+\alpha_3-1)\alpha_1^{\Dt-d_1-1}\alpha_2^{\Dt-d_2-1}\alpha_3^{\Dt-d_3-1}}\]
\be  \label{FT}
\tcboxmath{\times \frac{\Gamma(D-d_1-d_2-d_3)}{(k_1^2 \alpha_2\alpha_3+k_2^2 \alpha_1\alpha_3+k_3^2 \alpha_2\alpha_1)^{D-d_1-d_2-d_3}}}
\ee
Note that when $d_1+d_2+d_3=D$, the $k$ dependence disappears, and the answer is divergent. 

For the problem at hand $d_i=1$. So we get a factor  
$\Gamma(D-3)$.

\end{appendices}

\end{document}